\documentclass[twocolumn]{aastex631}

\usepackage{siunitx}
\usepackage{bm}
\usepackage[normalem]{ulem}

\usepackage{mathtools}
\usepackage{multirow}
\usepackage{hyperref}

\defcitealias{Secunda_2019}{Paper~I}
\defcitealias{Secunda_2020}{Paper~II}

\usepackage{graphicx}	

\begin{document}
\title[Hierarchical Black Hole Merger Mass Distribution]{Orbital Migration of Interacting Stellar Mass Black Holes in Disks around Supermassive Black Holes. III. Mass Distribution of Hierarchical Mergers}

\author[0009-0005-5615-760X]{Katherine L. Gonglewski}\thanks{E-mail:  klg3@arizona.edu}
\affil{Lunar and Planetary Laboratory, University of Arizona, Tucson, AZ, 85721, USA}

\author[0000-0002-1174-2873]{Amy Secunda}
\affil{Center for Computational Astrophysics, Flatiron Institute, New York, NY 10010, USA}

\author[0000-0003-0064-4060]{Mordecai-Mark Mac Low}
\affil{Department of Astrophysics, American Museum of Natural History, 200 Central Park West, New York, NY, 10024, USA}

\author[0000-0002-5956-851X]{K. E. Saavik Ford}
\affil{Department of Astrophysics, American Museum of Natural History, 200 Central Park West, New York, NY, 10024, USA}
\affil{Department of Science, CUNY Borough of Manhattan Community College, 199 Chambers Street, New York, NY 10007, USA}
\affil{Physics Program, CUNY Graduate Center, 365 5th Avenue, New York, NY 10016, USA}

\author[0000-0002-9726-0508]{Barry McKernan}
\affil{Department of Astrophysics, American Museum of Natural History, 200 Central Park West, New York, NY, 10024, USA}
\affil{Department of Science, CUNY Borough of Manhattan Community College, 199 Chambers Street, New York, NY 10007, USA}
\affil{Physics Program, CUNY Graduate Center, 365 5th Avenue, New York, NY 10016, USA}

\author[0000-0002-5965-1022]{Fabian R. N. Schneider}
\affil{Heidelberg Institute for Theoretical Studies, Schloss-Wolfsbrunnenweg 35,
69118 Heidelberg, Germany}
\affil{Zentrum f{\"u}r Astronomie der Universit{\"a}t Heidelberg, Astronomisches Rechen-Institut, M{\"o}nchhofstr.\ 12-14, 69120 Heidelberg, Germany}




\begin{abstract}
Active galactic nucleus (AGN) disks are a promising location for the formation of binary black holes (BBHs) that will merge on relatively short timescales and be detected by LIGO-Virgo-KAGRA (LVK). To compare the mass function (MF) of black holes (BHs) undergoing hierarchical mergers in AGN disks to the inferred MFs from LVK observations, we perform 360 simulations with an N-body code augmented to include an analytic model for migration torques and other gas forces. We focus on the region surrounding migration traps in AGN disks where migration torques cancel out and BHs converge. We find that regardless of changes in the initial MF and BBH merger criteria, frequent mergers deplete the number of BHs with masses $\lesssim 10$~$M_\odot$ and fill the upper mass gap with a roughly uniform distribution from 40--100~$M_\odot$, with a slight overabundance around ${\approx}\,70\,M_\odot$ from resonant orbiters. We also find an average merger rate of $\sim 6$~Gpc$^{-3}$~yr$^{-1}$ for migration-trap-aided BBH mergers in our AGN disk model. $\sim 40\%$ of these mergers have uneven mass ratios and 16\% have a primary mass $\in[50-100]~M_\odot$. Therefore, AGN disks could easily be the source of BBH mergers observed by LVK that are difficult to produce through traditional stellar evolution channels. Our simulations also form a separate higher-mass intermediate mass black hole (IMBH) population $>200~M_\odot$ after $\sim 2$~Myr. Future gravitational wave detectors can use observations of this IMBH population to constrain models of AGN accretion disks.


\end{abstract}




\section{Introduction}
LIGO-Virgo-Kagra (LVK) has detected over 150 binary black hole (BBH) mergers through gravitational waves \citep{GWTC4a}. The details of these mergers---the mass of the primary, the mass ratio between the primary and the secondary, and the effective spin of the binary, as well as the correlations between these parameters---provide important clues to their astrophysical origins. For example, over $20\%$ of the BBH mergers in the fourth LVK Gravitational-Wave Transient Catalog (GWTC-4) involved at least one black hole (BH) that fell within the upper mass gap between $\sim $40--120~$M_{\odot}$ \citep{GWTC4a,GWTC4}, where it is challenging to form BHs via traditional models of stellar evolution \citep{Heger_2002,Woosley_2007,Farmer_2019,Woosley_2021,Renzo_2024,Broekgaarden_2022,VanSon_2020}. Instead, hierarchical mergers, where one or both components of a BBH merger are already the product of one or more BBH mergers, may account for this high mass population of BHs \citep[e.g.,][]{Gerosa21}.

Hierarchical mergers are expected to occur in various types of stellar clusters \citep[e.g.][]{Mapelli21}, such as young stellar clusters \citep[e.g.][]{Kremer20}, nuclear star clusters \citep[e.g.][]{Antonini16,Fragione19}, and globular clusters \citep[e.g.][]{Rodriguez18,Kremer20}. Hierarchical mergers could also occur in active galactic nucleus (AGN) disks \citep[e.g.][]{McK14,Bellovary_2016,Bartos17,Stone17,Secunda_2019,Secunda_2020,Tagawa20,FM25,Moncrieff26}. AGN disks are particularly intriguing locations for producing hierarchical mergers because the gas disk can exchange angular momentum with the orbiters driving orbital migration inwards \citep{Goldreich1979,Ward1997,Tanaka2004} and in some cases outwards \citep{Paardekooper2006,Paardekooper_2010}. In addition, the gas disk dampens the inclination and eccentricity of embedded orbiters.

\cite{Bellovary_2016} showed that for both \cite{Sirko&Goodman_2003} and \cite{Thompson2005} AGN disk models, there is a region of the AGN disk where the outward and inward migration torque cancels out, called a migration trap. With N-body simulations, \citet[][hereafter \citetalias{Secunda_2019}]{Secunda_2019} showed that BHs rapidly migrate towards these traps facilitating frequent close encounters at low relative velocities, significantly increasing the likelihood of hierarchical mergers. As a result, migration traps greatly facilitate the formation of BBHs in the upper mass gap and intermediate mass black holes \citep[IMBHs; e.g. \citetalias{Secunda_2019};][]{McK12,Bellovary_2016,Yang19,Tagawa21, PengXian21,Vaccaro25,Secunda_2020}. Monte Carlo simulations suggest that migration ``swamps'' or ``traffic jam'' regions of the disk where migration torques change significantly in magnitude (without changing sign) also have rapid hierarchical mergers, perhaps even at a higher rate than migration traps \citep{Vaccaro25,McKernan:2025}. 

The AGN disk may also be an ideal environment for generating BBHs that match other properties observed by LVK. For example, the population of BBH mergers occurring within an AGN disk may have a bias in the effective spin parameter, $\chi_{\rm eff}$, towards positive values due to symmetry-breaking effects \citep[e.g.][]{Wang21}. In addition, Monte Carlo simulations show that, depending on model assumptions, mergers in an AGN disk can also produce the observed anti-correlation reported by \cite{Callister21} and \cite{Adamcewicz23} between the mass ratio $q$ and $\chi_{\rm eff}$ \citep{McK22,Santini23, Vaccaro23,Cook:2025,Su_2025,Delfavero25}. 

Determining the population statistics for different astrophysical BBH formation channels is crucial to understand the influence these channels have on the observed LVK population. Until now, population studies of the AGN merger channel have been limited to examinations using Monte Carlo methods or 1-D N-body simulations \citep[e.g.][]{Yang19,McK20,Tagawa20,Tagawa21,Gilbaum:2025,Rowan:2025, McKernan:2025, Cook:2025,Tagawa26}. In this paper, we perform a population study with hundreds of simulations using a 3-D N-body code that has been augmented to include forces from the AGN gas disk. The advantage of using an augmented 3-D N-body code is that it explicitly integrates the phase-space evolution of the orbiters under the influence of gravitational and gas forces with a resolved timestep, as opposed to Monte Carlo methods which evolve orbiters statistically with a large timestep. However, because N-body codes are significantly more computationally expensive, Monte Carlo codes are able to perform wider parameter studies. Therefore, 3-D N-body simulations are complementary to Monte Carlo and 1-D simulations, and the results from these different methods should be compared. 

We initialize our N-body simulations with two different initial mass functions (IMFs): the empirical \citet{Salpeter_1955} IMF for stellar populations and a bimodal BH IMF predicted by \citet{Schneider_2023} from models of isolated binary stellar evolution. We also test the assumptions of the N-body simulations in \citetalias{Secunda_2019} and \citetalias{Secunda_2020} by applying different BBH formation and merger criteria based on recent hydrodynamical simulations of BBHs embedded in gas disks \citep{Rowan2023,Qian2024,Whitehead2024,DeLaurentiis2023,Whitehead25}. We then compare the resulting mass distributions of our hundreds of simulations to the mass function (MF) inferred from GWTC-4 \citep{GWTC4}. This comparison helps us constrain the role migration-assisted mergers in an AGN disk play in forming the GWTC-4 MF and the disk lifetime required to build up the observed MF in the mass gap. We also make predictions for IMBH formation, which will eventually be testable with LISA, and will help to constrain properties of AGN disks.

\section{Methods}

In order to compare the MF of BHs formed through mergers in an AGN disk to the inferred GWTC-4 MF \citep{GWTC4} for different BH IMFs and merger criteria, we perform 360 simulations with the augmented N-body code from \citetalias{Secunda_2019} and \citetalias{Secunda_2020}. We run each simulation for 10~Myr, a rough upper estimate for the lifetime of an AGN disk \citep[e.g.,][]{Haehnelt:1993,Marconi:2004}. In \S~\ref{sec:nbody} we briefly describe our augmented N-body code, and in \S~\ref{sec:imf_choices} we describe the initial parameters for the orbiters in our simulations, including details on our two different choices for the BH IMF. 

\subsection{N-body Code}
\label{sec:nbody}

Our simulations are performed using a Bulirsch-Stoer N-body code augmented to include gas forces. This code was originally used to study protoplanetary migration \citep{Sandor_2011, Horn_2012} and was adapted in \citetalias{Secunda_2019} to simulate the migration of stellar-mass black holes in AGN disks. The gas forces added to the N-body integration include eccentricity and inclination dampening given by \cite{cresswell2008}, turbulent density fluctuations following \cite{Laughlin_1994} and \cite{Ogihara_2007}, and most importantly, migration torques determined by the analytic prescription in \citet{Paardekooper_2010}, which depends on the local gradients of density, temperature, and entropy, as well as the optical depth. As in \citetalias{Secunda_2019}, the properties of the AGN gas disk are modeled after \citet{Sirko&Goodman_2003} with a central supermassive black hole (SMBH) with mass $M_{\rm SMBH}=10^{8}\, M_\odot$, an integrated disk mass out to $2 \times 10^{5}~R_g$ of $3.7 \times 10^{7}\, M_{\odot}$, and an Eddington accretion ratio fixed at 0.5. Using the Monte Carlo code \texttt{McFACTS}, \citet{Delfavero25} found that BBH mergers occur with the highest frequency in AGN disks around SMBHs with masses close to $10^{8}\, M_\odot$. However, in the future, testing a wider range of SMBH masses and accretion rates with our N-body code will provide a more complete picture.

In our N-body simulation we focus on the region from 200--1000~$R_g$ directly surrounding the migration trap at  $\sim 650$~$R_g$. We note that the location of the migration trap in our simulations is specific to the parameters of our disk model. The exact locations where migration traps occur in disk models, and whether they occur at all \citep{Pan21}, can vary when thermal feedback is included in torque calculations \citep[e.g.][]{Grishin24,Gilbaum:2025}. 

In \citetalias{Secunda_2020}, we included additional orbiters that migrate inwards at the outer boundary to represent both BHs that migrate in from the outer disk and initially highly inclined orbiters that are ground down into alignment with the disk \citep{Fabj20,Nasim23,WZL2024,Rowan:2025}. We assumed that there are roughly 10 BHs at any given time within a few thousand gravitational radii beyond our simulated disk or with inclinations $\lesssim20^\circ$, for which both the migration timescale and the grind-down timescale for disk capture are $\sim 10^6$~yr. These assumptions give us a rate of one additional BH at the outer boundary every 100~kyr. For simplicity in \citetalias{Secunda_2020} the masses of these BHs were set uniformly to 10~$M_{\odot}$. Here, we draw the masses of these additional orbiters from the IMF of each simulation, which allows for a more realistic distribution of BH masses throughout the simulation. However, more massive BHs will both migrate and be ground down into alignment with the disk more rapidly. Therefore, future work could increase the sophistication of our simulations further by increasing the rate of more massive incoming orbiters relative to less massive incoming orbiters.

Due to computational limits, as in \citetalias{Secunda_2019}, two orbiters are considered merged as soon as they form a BBH. The conditions for forming a BBH are: 1) the relative kinetic energy of the binary components is less than the binding energy of the BBH and 2) the distance between the BBH components  $\Delta R \leq 1 R_{\rm H}$, where $R_{\rm H}=a (Q/3)^{1/3}$ is the mutual Hill radius of the BBH, $a$ is the semi-major axis of the binary center of mass, and $Q=M_{\rm bin}/M_{\rm SMBH}$ is the mass ratio of the binary to the SMBH. Recent studies have explored the role of gas-assisted dynamical friction in further facilitating BBH mergers in AGN disks. Notably, 2D hydrodynamic simulations \citep[e.g.][]{YaPingLi21,RixinLi22,Jairu23} have begun to reveal a fractal parameter space \citep{Rowan2023,Qian2024,Whitehead2024,DeLaurentiis2023} suggesting that BBH binding occurs at larger radii than our default assumption, since otherwise scattering can cause unbinding. Recent 3-D simulations seem to confirm this result \citep{Whitehead25}. To test the effect of these findings on our N-body simulations, we perform two additional case studies (using the Salpeter IMF) where instead of requiring $\Delta R \leq 1 R_{\rm H}$, we require $\Delta R \leq 0.7 R_{\rm H}$ or $\Delta R \leq 1.5 R_{\rm H}$. 

\begin{figure}
\includegraphics[width=\columnwidth]{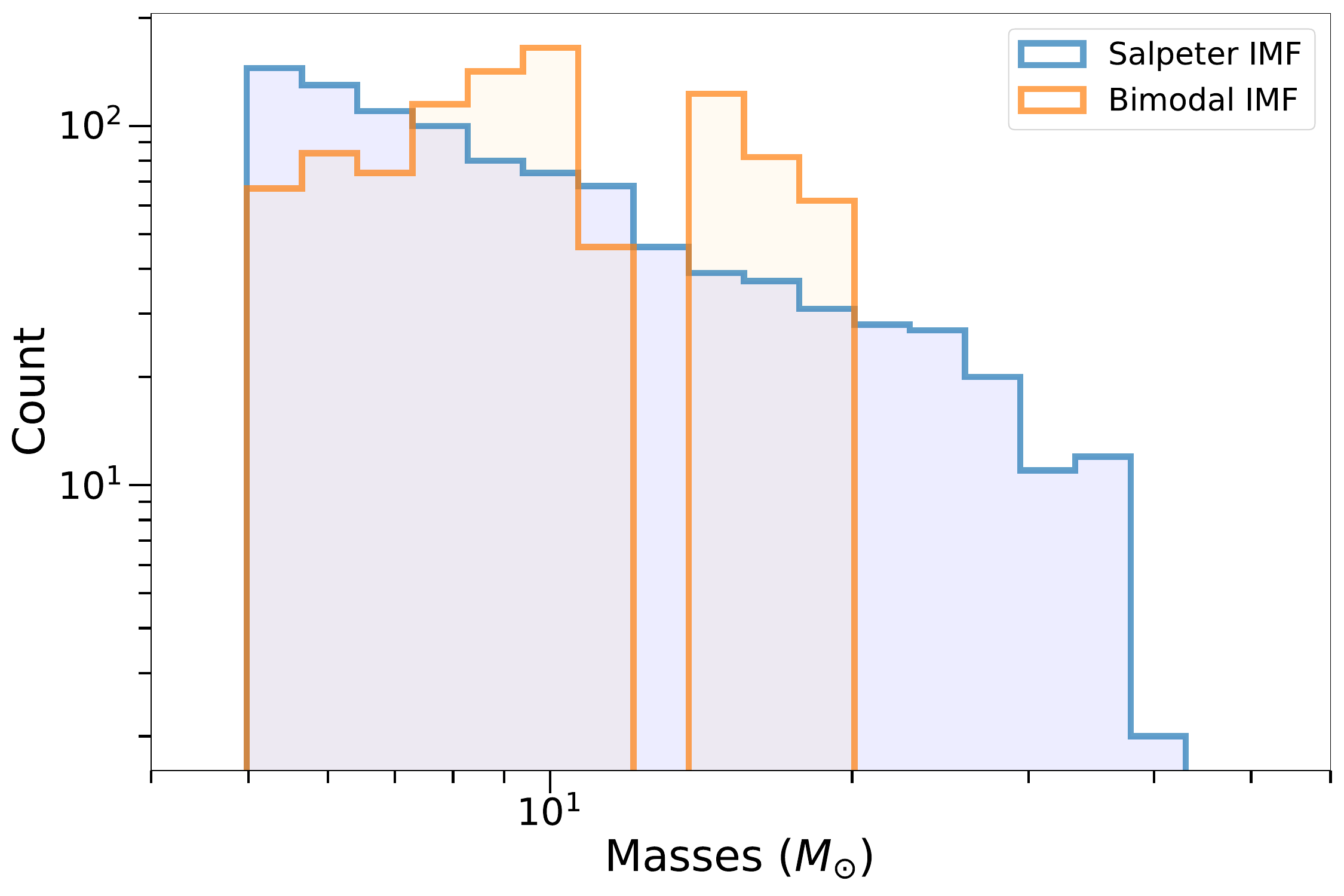}
    \caption{The two different BH IMFs used in our simulations, overlaid for ease of comparison. Note that due to the imposed mass cut-off the highest bin centered at 40~$M_{\sun}$ does not contain any BHs above 40~$M_{\sun}$.}
    \label{fig:3IMFs}
\end{figure}

\begin{figure*}
\includegraphics[width=\textwidth]{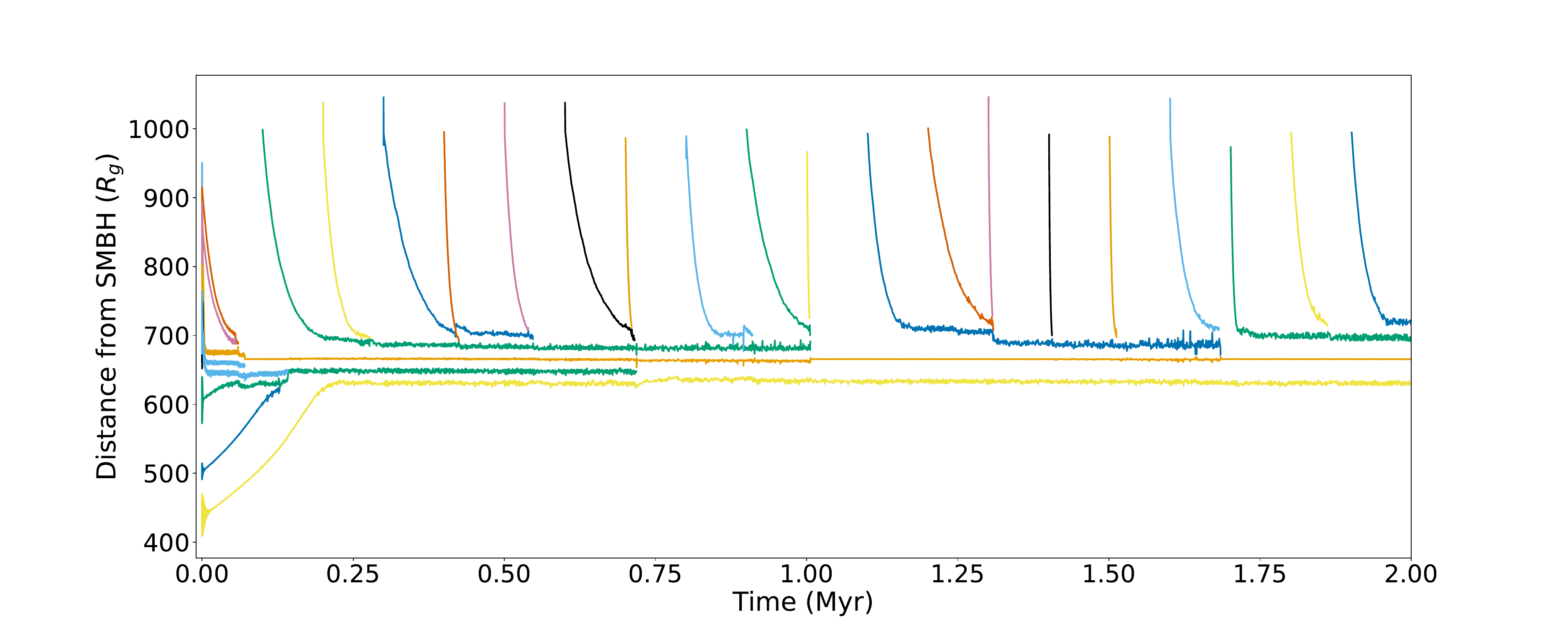}
        \caption{The distance of BHs from the SMBH (in units of the gravitational radius, $R_g$) as a function of time for the first 2~Myr of a single example Salpeter IMF simulation. Each colored line represents a different BH, with only the more massive continuing after a merger has occurred. We see several examples of BHs ending up in resonant orbits with the BH in the migration trap at $\sim 650$~$R_{g}$ and numerous BBH mergers. Resonant orbiters outside the migration trap are disrupted by newly arriving BHs introduced at the outer boundary of the simulated disk, but inner resonant orbiters (such as the yellow BH) can remain stable over $>2$~Myr timescales.
    } 
    \label{fig:singlerun}
\end{figure*}

\subsection{Initial Mass Function and Orbiter Parameters}
\label{sec:imf_choices}
We test the impact of two different BH IMF choices on the final MFs of BHs in our simulations. For our default IMF, we use the \cite{Salpeter_1955} MF inferred from stellar populations with $dN/dm \propto M^{-2.35}$. We draw $10^6$ random values from this distribution using a minimum mass of $5~M_\odot$ and an upper mass cut-off at $40~M_\odot$. From these $10^6$ values, we randomly draw ten values that we use to initialize the masses of ten BHs at the start of each simulation. We also randomly draw values from this distribution for the masses of the additional migrating bodies that are added at the outer boundary every $10^5$~yr. 

In addition to this empirical IMF inferred from stellar populations, we use a second physically motivated BH IMF from \cite{Schneider_2023}. \cite{Schneider_2023} predict the mass distribution of field BHs formed from binary-stripped stars using stellar evolution models evolved with {\sc mesa} and a Salpeter IMF for the primary stars and single stars. They find that binary-stripped stars give rise to a bimodal BH mass spectrum with characteristic BH masses of about 9~$M_{\odot}$ and 16~$M_{\odot}$ due to carbon and neon burning becoming neutrino dominated. Here, we apply their solar metallicity model such that there are no BH masses beyond $\sim\,20\,M_{\odot}$, but this mass spectrum could extend up to over $40\,M_\odot$ for lower metallicities. As with the Salpeter IMF simulations, we randomly draw masses from this bimodal IMF for the ten BHs at the start of each simulation and then randomly draw additional masses for the BHs we add at the outer boundary during the simulation.

We show the Salpeter and bimodal IMFs in Figure \ref{fig:3IMFs}. We restrict both IMFs to 5~$M_{\odot}<M<40$~$M_{\odot}$. Our minimum mass cut-off is similar to the truncation in the LVK MF. We use our upper mass cut-off, at the bottom edge of the upper mass gap, to test whether hierarchical mergers in our AGN disk simulation lead to merging BBHs in this mass gap.

For all simulations, the other initial parameters of our orbiters are chosen as in \citetalias{Secunda_2019} and \citetalias{Secunda_2020}. The starting inclination and eccentricity are drawn from random Gaussian distributions with means of 0.05 and 0, and standard deviations of 0.02 and 0.05, respectively. The absolute value of the inclination is used, while for the eccentricity if the randomly generated value is negative, we draw a new value. The initial phases, arguments of pericenter, and lines of nodes are chosen from a uniform random distribution. The semi-major axes of the initial ten orbiters in our simulation are chosen from a uniform random distribution between 200--1000~$R_g$.

\section{Results}

\begin{figure*}
\includegraphics[width=\columnwidth]{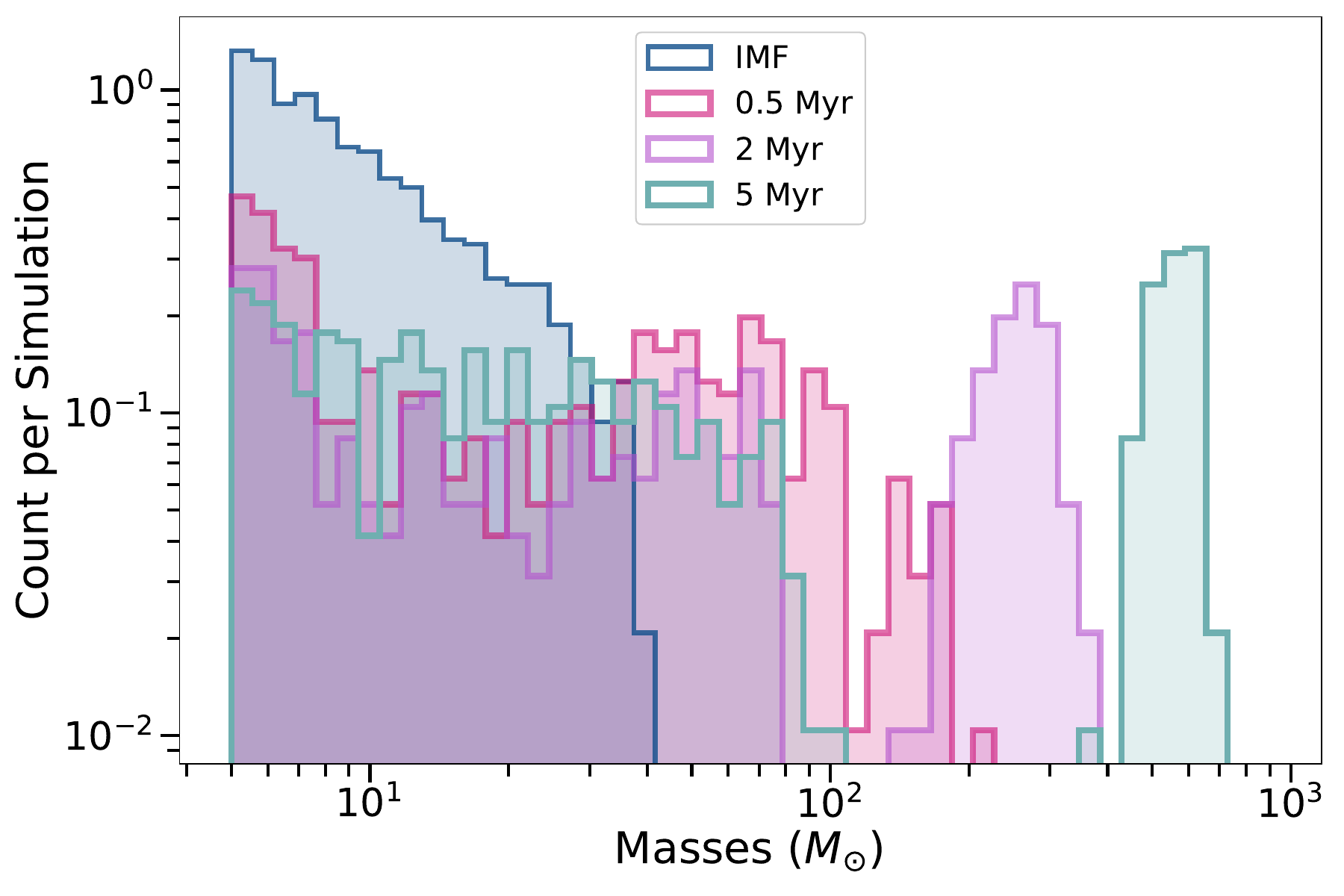}
\includegraphics[width=\columnwidth]{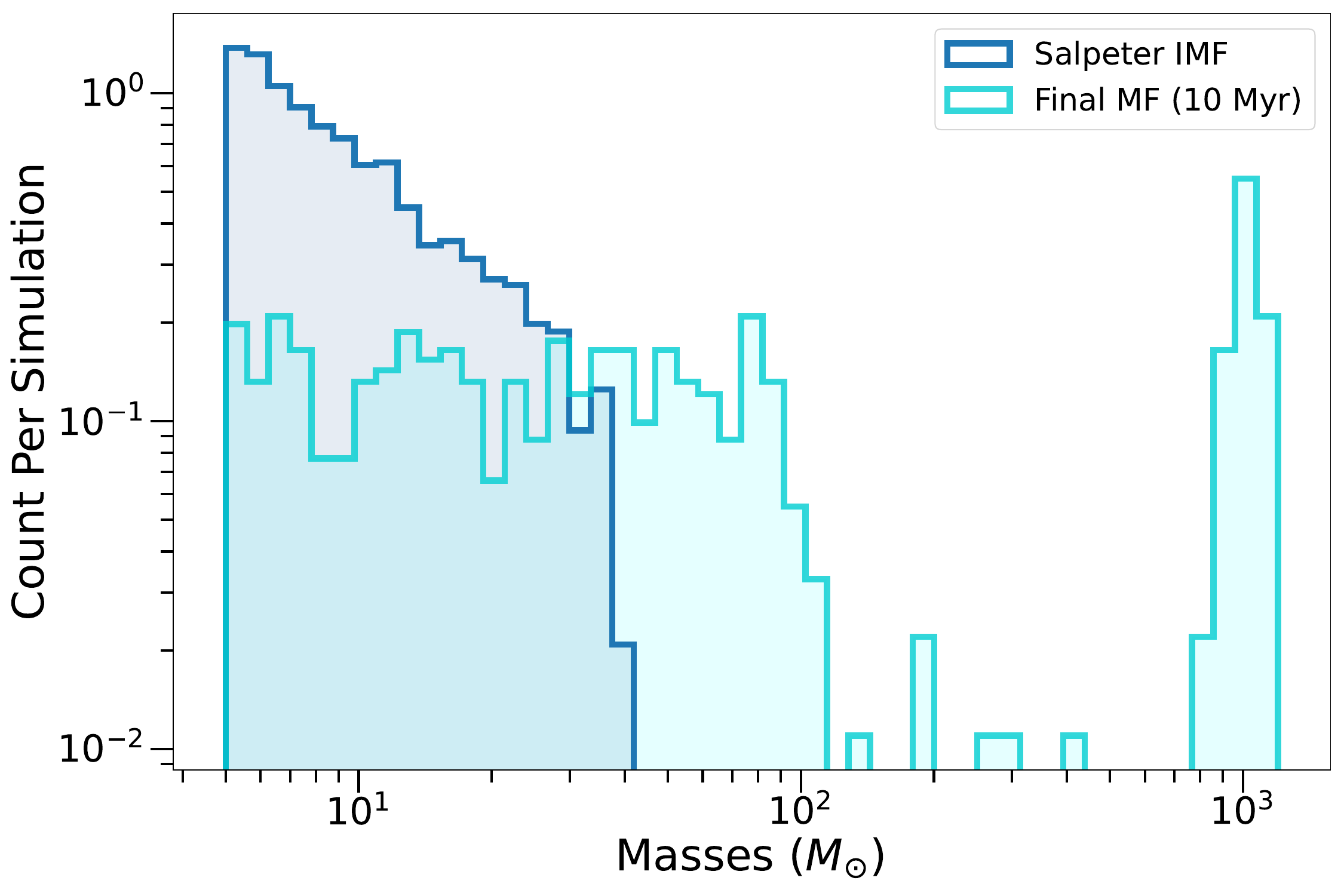}
    \caption{The left panel compares the IMF {\em (dark blue)} with MFs after 0.5~Myr (\emph{red}), 2~Myr (\emph{pink}), and 5~Myr (\emph{teal}) averaged over the 90 Salpeter simulations. The right panel compares the initial {\em (dark blue)} and final {\em (light blue)} MFs after 10 Myr averaged over the 90 Salpeter simulations.}
    \label{fig:IMFalternates}
\end{figure*}

\subsection{Salpeter IMF Runs}

\subsubsection{Evolution Over Time}
\label{sec:example}
We first examine the evolution of our Salpeter simulations over their 10~Myr durations. Figure \ref{fig:singlerun} shows the first 2~Myr of a typical individual simulation. It contains several common features that occur in a majority of our simulations, regardless of IMF. Our simulations typically begin with rapid first, second, and third-generation mergers of lower mass BHs. After less than 0.25~Myr the most massive BH tends to end up on a stable orbit in the migration trap. In this simulation two other BHs form a resonant chain with the BH in the migration trap. Because these BHs orbit at smaller radii than the migration trap radius, they typically do not interact with BHs farther out than the migration trap. As a result, these resonant orbiters remain stable for 0.5~Myr (green BH) and 3~Myr (yellow BH), when they are eventually disrupted by multi-body scatterings.

After the most massive BH ends up in the migration trap, the next BH that migrates inward from the outer disk is typically trapped in a first-order ($k+1:k$, where $k$ is the resonant harmonic) resonant orbit with the BH in the migration trap, $\sim30R_g$ outside of the migration trap. As additional incoming BHs migrate towards the trap, they most commonly merge with the BH in this resonant orbit increasing its mass over time. In addition, after each of these mergers the resonant orbiter moves inward towards the migration trap to a new resonant orbit with a higher harmonic. When this BH has undergone enough mergers to grow to a mass of $\sim 80~M_\odot$, it will have a high enough harmonic to reach resonant overlap \citep{Wisdom1980}, leading it to merge with the BH in the migration trap. Then, the next incoming BH will end up on a resonant orbit, and the pattern will repeat. 

From this pattern, we can infer that the population of BHs will fall into three main categories:
\begin{enumerate}
    \item recent incoming orbiters that have not undergone a first-generation merger and will have a mass drawn from the IMF,
    \item the BH in the migration trap, which grows to a mass roughly equal to the total mass of all orbiters in the simulation ($\sim 1000\,M_{\odot}$ after 10~Myr),
    \item one to three resonant orbiters that have undergone at least one merger and will over time accumulate in mass through additional mergers from $\sim 15 -80~M_{\odot}$.
\end{enumerate}

We show how these main categories build up over time at the population level across our 90 Salpeter simulations in Figure \ref{fig:IMFalternates}. The left panel compares the IMF averaged over our 90 Salpeter simulations with the average MF after 0.5~Myr, 2~Myr and 5~Myr, and the right panel compares the initial and final MF after 10~Myr averaged over our 90 Salpeter simulations. After only 0.5~Myr, the Salpeter IMF is still apparent in the peak of low mass BHs and the dearth of BHs around 10--30~$M_\odot$. However, hierarchical mergers have already occurred, creating a peak from $30-100~M_\odot$ and filling the upper mass gap up to $\sim 200~M_{\odot}$. Around 2~Myr there are fewer low-mass BHs below $10~M_\odot$, and the small peak from hierarchical mergers tightens to between $30-80~M_\odot$. In addition, a distinct IMBH population between $150-400~M_\odot$ now breaks off from the main BH MF. This IMBH mass gap widens with time as the mean mass of the massive IMBH distribution continues to grow steadily. By 5~Myr, this IMBH peak extends to over $700$~$M_{\odot}$. 

At 5~Myr, the MF below $80~M_\odot$ has flattened. Over the next 5~Myr, the MF will continue to flatten the slight remaining power-law slope from the Salpeter IMF, because the low mass BH population ($<10$~$M_{\odot}$) merges to form higher mass BHs at a faster rate than it is replenished by migration and disk capture. This rapid merger rate leads to a drop in the MF around 10~$M_{\odot}$, because our mass cut-off at 5~$M_{\odot}$ means that after an initial merger all BHs will have a mass greater than 10~$M_{\odot}$. Instead, there is a small peak from $\sim$ 10--20~$M_{\odot}$ that is not native to our IMF made up of the products of low mass first-generation mergers.

In our final MF, hierarchical mergers fill the upper mass gap ($>40$~$M_{\odot}$) up to $\lesssim 100$~$M_{\odot}$. The MF between 15--80~$M_{\odot}$ is made up of BHs in resonant orbits with the BH in the migration trap that have undergone at least one merger and will over time build up in mass through subsequent mergers before merging with the BH in the migration trap (category 3 above). Because the 10~Myr cut-off can occur at any random point in the growth of these resonant orbiters, we have a roughly uniform distrib ution of the masses between their initial mass of $\sim 15$~$M_{\odot}$ and the final mass at which they frequently merge with the BH in the migration trap, $\sim 80~M_{\odot}$. In this mass range there are two small peaks, one around $35~M_\odot$ and a second at the top end of the range around $60-80~M_{\odot}$. While the hierarchical mergers in our simulations fill the upper mass gap to $\lesssim 100$~$M_{\odot}$, a new mass gap forms from $\sim 100-800~M_{\odot}$, because resonant orbiters rarely reach $\gtrsim100~M_{\odot}$ without merging with the BH in the migration trap. Above $800~M_{\odot}$ a massive IMBH peak made up of BHs in migration traps (category 2 above) forms at $1000\pm100~M_\odot$. Thus, our simulations drive efficient oligarchic growth of IMBHs up to $\sim 10^{3}$~$M_{\odot}$ in $\leq 10$~Myr.


\begin{figure}
    \centering
    \includegraphics[width=\columnwidth]{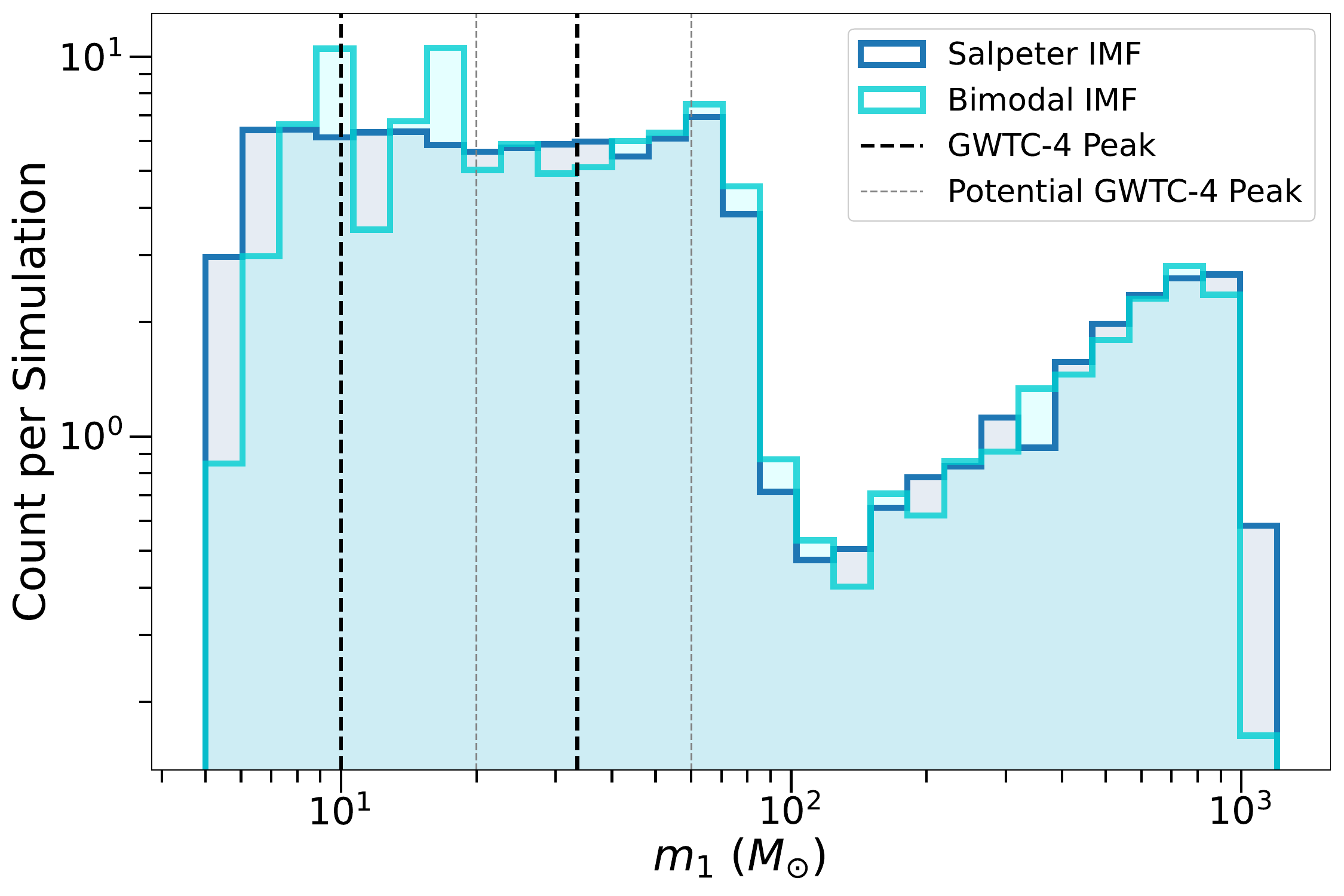}
    \caption{The average number of BBH mergers per simulation as a function of the primary mass for all Salpeter simulations (\emph{dark blue}) and all bimodal simulations (\emph{light blue}). For comparison, dashed lines show the location of peaks reported in GWTC-4 for the merging MF.}
    \label{fig:primary_log}
\end{figure}

\begin{figure}
\includegraphics[width=\columnwidth]{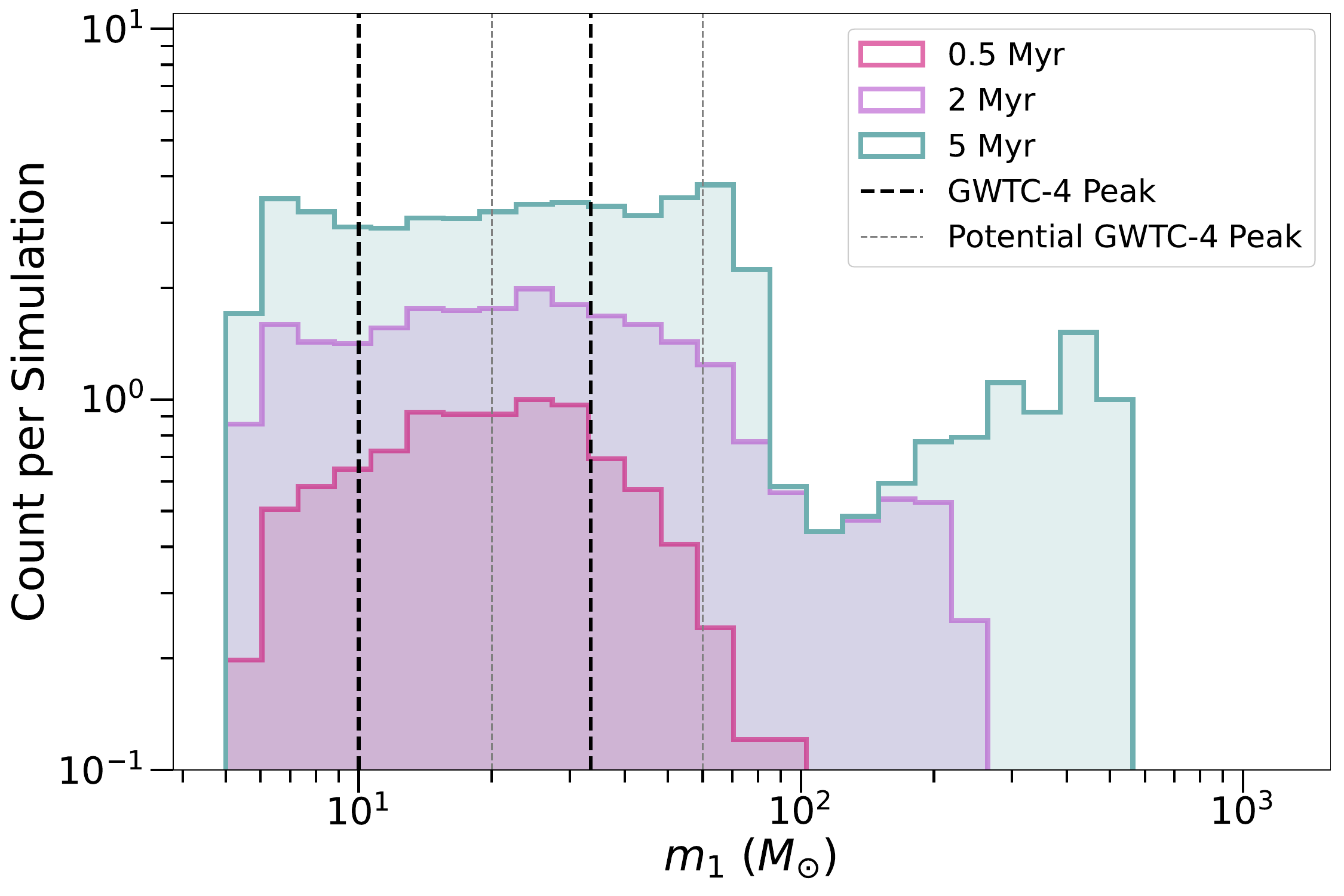}
   \caption{The average number of BBH mergers per simulation for all Salpeter simulations as a function of the primary mass for disk lifetimes of 0.5, 2, and 5~Myr. For comparison, dashed lines show the location of peaks reported in GWTC-4 for the merging MF.}
    \label{fig:evolutionhist}
\end{figure}

\begin{figure*}
\includegraphics[width=0.49\textwidth]{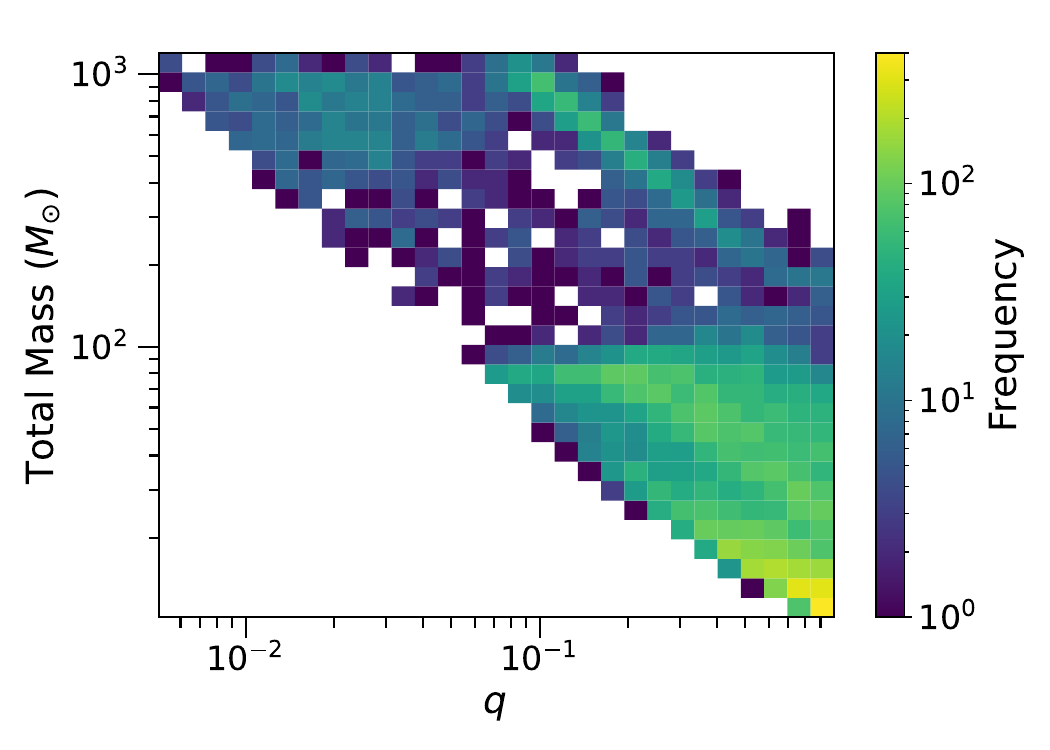}
\includegraphics[width=0.49\textwidth]{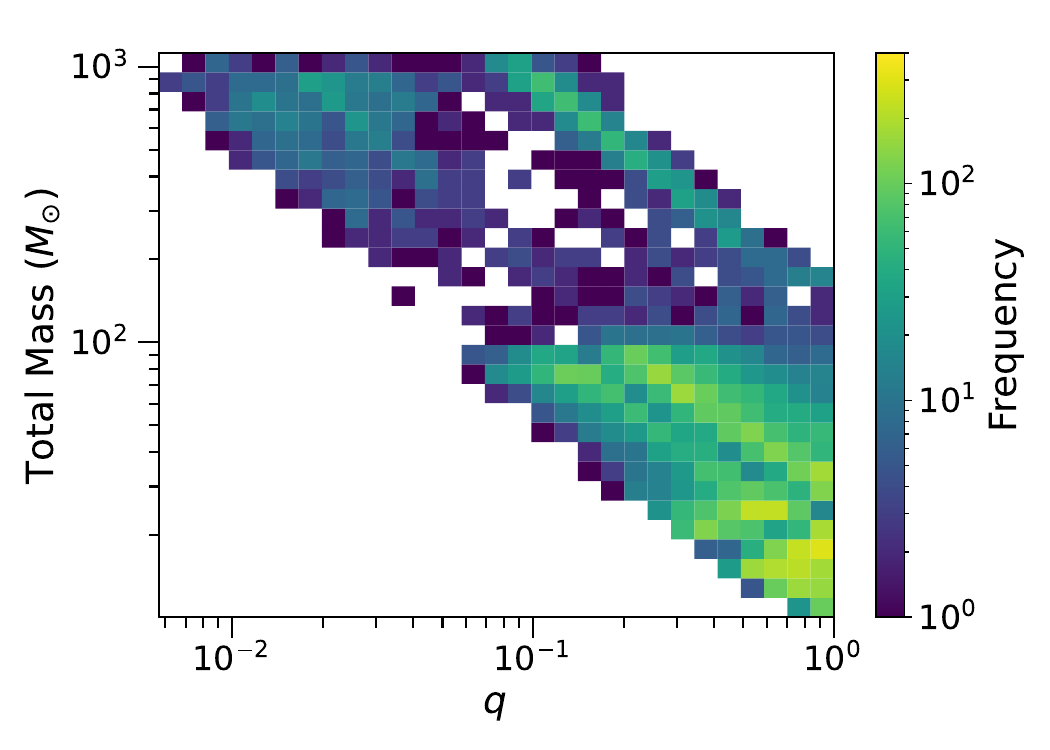} \\
\includegraphics[width=0.49\textwidth]{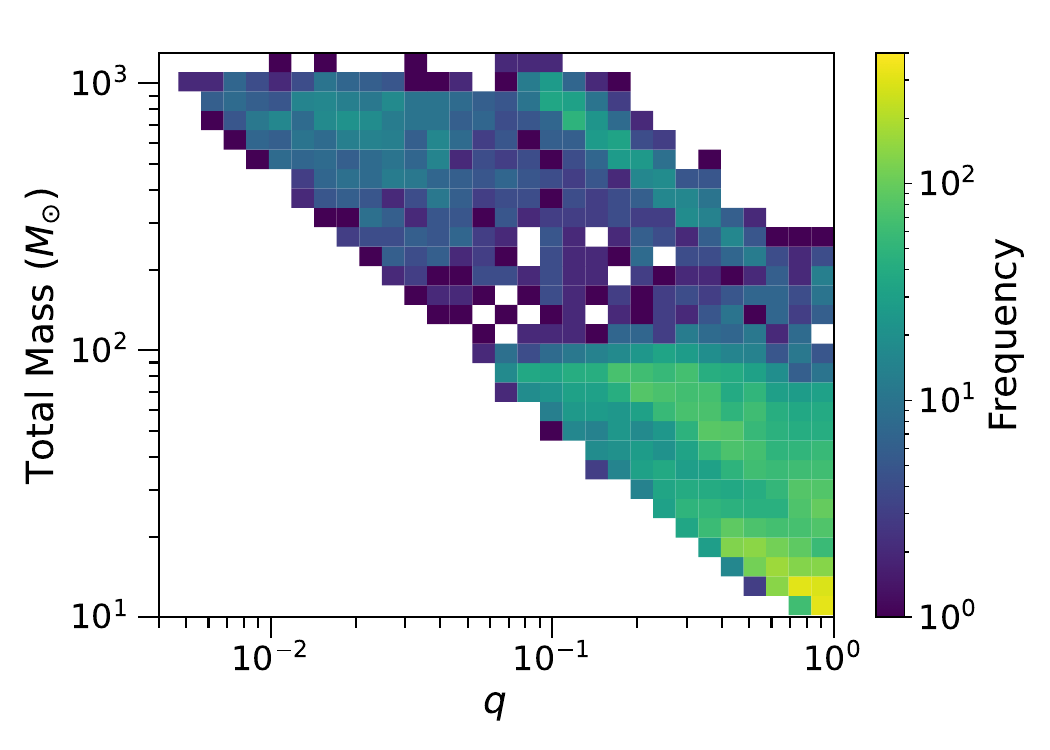}
\includegraphics[width=0.49\textwidth]{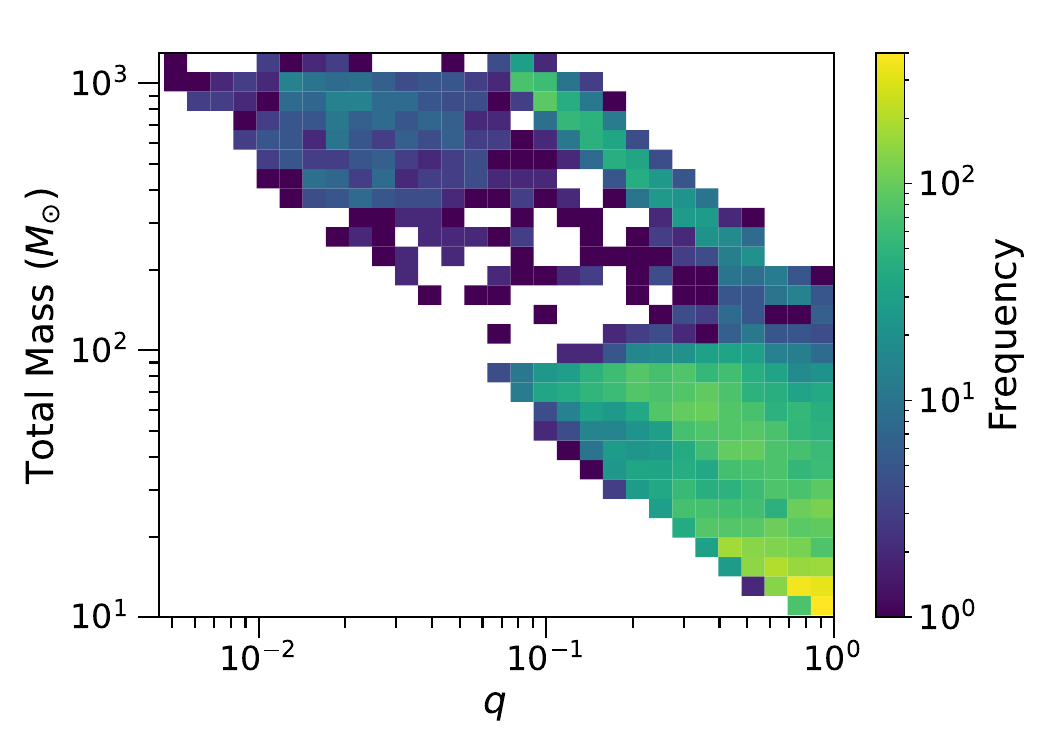}
    \caption{Frequency of BBH mergers from our 90 Salpeter {\em (top left)}, bimodal {\em (top right)}, $\Delta R=0.7~R_H$ {\em (bottom left)}, and $\Delta R=1.5~R_H$ {\em (bottom right)} simulations as a function of mass ratio $q=m_{2}/m_{1} \leq 1$ and total binary mass $M_{\rm bin}=m_{1}+m_{2}$.}
    \label{fig:qvMass_histogram}
\end{figure*}

\subsubsection{Merger Masses and Mass Ratios}
\label{sec:mq}

In addition to the evolution of the mass distribution of BHs in our Salpeter simulations, we can also examine the properties of the BBH mergers in these simulations. Figure \ref{fig:primary_log} shows the average distribution of primary masses for BBHs produced in our 90 Salpeter simulations in dark blue. For comparison the vertical dashed lines show masses at which the GWTC-4 primary MF peaks. We discuss this comparison in further detail in \S~\ref{sec:compare_gwtc3}. 

The greatest number of BBHs form with primary masses between $m_1 \sim5$--$70~M_{\odot}$. The lower limit of this mass range is set by our minimum IMF mass of $m_1 = 5~M_\odot$. The upper limit comes from the mass of the resonant orbiter when it undergoes its final merger with a lower mass inward migrating BH at a mass of $m_1 \approx 70~M_\odot$. After this final merger, the BH reaches the mass threshold we find for resonant orbiters in our simulations and merges with the BH in the migration trap as the secondary mass, $m_2 \approx80~M_\odot$. Because of this mass threshold for resonant orbiters, there is a dearth of primary masses between $\sim 70$--$300~M_\odot$, although on average we still expect a rate of $\sim3$ per AGN in this mass range. Our simulations also predict several mergers with $m_1>300~M_\odot$ per AGN due to a second peak in the distribution around $800~M_\odot$. The most massive primary masses in our Salpeter simulations exceed $1000~M_\odot$.

The lifetime of an AGN disk is uncertain and could be as short as 0.1~Myr \citep[e.g.,][]{Schawinski:2015,King:2015}. Therefore, in Figure~\ref{fig:evolutionhist}, we show what the primary MF would look like over shorter timescales.  If the lifetime of an AGN disk is only 0.5~Myr, the number of mergers per AGN disk would be much lower and the primary mass distribution would range from $5$--$100~M_\odot$, peaking around $20~M_\odot$. The number of mergers increases significantly as the lifetime of the disk increases, as does the range of the primary mass distribution. The peak in the primary mass distribution widens significantly for a disk lifetime of 2~Myr and includes a tail towards $m_1\gtrsim200~M_\odot$. As a result, an AGN with a disk lifetime of 2~Myr could easily account for the upper mass gap mergers observed by LIGO. By 5~Myr the peak has widened farther and a second peak around $400~M_\odot$ forms.

If the lifetime of the disk is $\lesssim 0.5$~Myr, migration-trap-assisted mergers would be unable to account for the upper mass gap mergers observed by LIGO, and IMBHs would only be present in AGN disks with lifetimes $\gtrsim5$~Myr. However, it is also possible that AGN with these shorter disk lifetimes would have episodic disk periods. If the gas disk appears episodically, then the higher masses of the final MFs at the end of each disk episode (see Figure \ref{fig:IMFalternates}) could produce larger BBH masses in each subsequent disk episode.

The top left panel of Figure~\ref{fig:qvMass_histogram} shows a two-dimensional histogram of the mass ratios $q=m_{2}/m_{1} \leq 1$ and total BBH masses $M_{\rm bin}=m_{1}+m_{2}$ for all mergers over 10~Myr for the 90 Salpeter simulations. Four over-densities are apparent. By an order of magnitude, the most common BBH properties in our simulation fall in the bottom right corner of of the histogram (high $q$, low $M_{\rm bin}$ mergers). The next most prominent over-density is a diagonal area in the lower right quadrant with $q > 0.1$ and $M_{\rm bin} < 100\, M_{\odot}$. There is also a narrow over-density along the upper right edge with $q > 0.1$ and $M_{\rm bin} = 10^{2}$--$10^{3}\,M_{\odot}$. Finally, there is a less prominent peak in BBH properties in the upper left corner with $q < 0.1$ and $M_{\rm bin} = 400$--1000~$M_\odot$.

As expected, the largest over-density is made up of first-generation encounters between low-mass BHs with similar masses ($q\sim 1$ and $M_{\rm bin} \sim 10~M_{\odot}$). We can also see instances of these first-generation mergers in Figure \ref{fig:singlerun}, mostly at the very start of the simulation or between two incoming BHs (e.g., the green and yellow lines at $\sim0.25$~Myr, the black and orange lines at 0.75~Myr, etc.). 

The next most frequent mergers, with $q \sim 0.1$--1 and $M_{\rm bin} <100\,M_{\odot}$, occur when lower mass incoming first- or second-generation BHs merge with a BH in a resonant orbit with the larger BH in the migration trap. These resonant orbiters grow from 10--80~$M_{\odot}$ over time, eventually leading to a lower mass ratio if they merge with an incoming BH that has undergone $<2$ mergers. We can see examples of these mergers in Figure \ref{fig:singlerun}, including the BHs merging with the blue orbiter from 1.25 through 1.75~Myr. Even mass, high total mass mergers are also possible if the incoming orbiter has undergone several mergers before merging with the resonant orbiter. An example of this in Figure \ref{fig:singlerun} is when the blue orbiter merges with the incoming orange and pink orbiters before merging with the green resonant orbiter at 0.6~Myr.

The third category, extends diagonally from $M_{\rm bin}\approx100$~$M_{\odot}$ to $M_{\rm bin}\approx1000$~$M_{\odot}$ $q\approx1$ down to $q = 0.1$, results from resonant orbiters merging with the higher mass BH in the migration trap. These mergers lead the BH in the migration trap to build up to a massive IMBH creating the peak around $\sim1000$~$M_{\odot}$ in the MF after $\sim 10$~Myr. Because resonant orbiters almost always merge with the BH in the migration trap once they reach a mass of $\sim80~M_{\odot}$, the highest mass mergers in this category will always have low mass ratios, leading to this clear diagonal over-density. In Figure \ref{fig:singlerun} this type of merger occurs around both 1 and 1.75~Myr when the green and then blue resonant orbiters merge with the orange orbiter in the migration trap.

A less prominent over-density of mergers in $M_{\rm bin}$-$q$ space occurs around $q \sim 0.01$--0.1 and $M_{\rm bin}>400\,M_{\odot}$. This last population corresponds to mergers directly between low-mass first- or second-generation BHs and the large IMBH growing in the migration trap. We can see in Figure \ref{fig:singlerun} that these mergers are uncommon, because the incoming BHs usually merge with a resonant orbiter first rather than directly with the IMBH in the migration trap. However, on rare occasions, these mergers do occur due to gas turbulence and multi-body interactions. For example, around 3~Myr into the simulation shown in Figure \ref{fig:singlerun}, the yellow orbiter that is on a stable resonant orbit at a smaller radius than the migration trap and has not merged with any other BHs becomes destabilized due to a multi-body interaction and merges with the BH in the migration trap. While uncommon, we point out that if a merger is observed with this high mass and low mass ratio it could come from an AGN disk channel (see \citetalias{Secunda_2020} for more details on these rare mergers) and not necessarily be a stellar mass BH merger with the IMBH at the center of a globular cluster.

There is also an under-density in $M_{\rm bin}$-$q$ space in our simulations around $q = 0.1$ and $M_{\rm bin} = 100$--300~$M_{\odot}$. This under-density is the result of the apparent maximum mass of the resonant orbiter which is just below $100\, M_{\odot}$. Due to this maximum mass, mergers with $M_{\rm bin} = 100$--300~$M_{\odot}$ will not occur between low mass first- or second-generation BHs and the resonant orbiter. Mergers between BHs in the migration trap and low mass first- or second-generation BHs at earlier times are also especially rare because the gravitational perturbations that lead to these already uncommon mergers will not be strong enough until later times when the BH in the migration trap is larger. If LVK detects mergers with $q = 0.1$ and $M_{\rm bin} = 100$--300~$M_{\odot}$ they are unlikely to originate in the migration trap region of an AGN disk.



\subsection{Bimodal IMF Runs}

\begin{figure}
\includegraphics[width=\columnwidth]{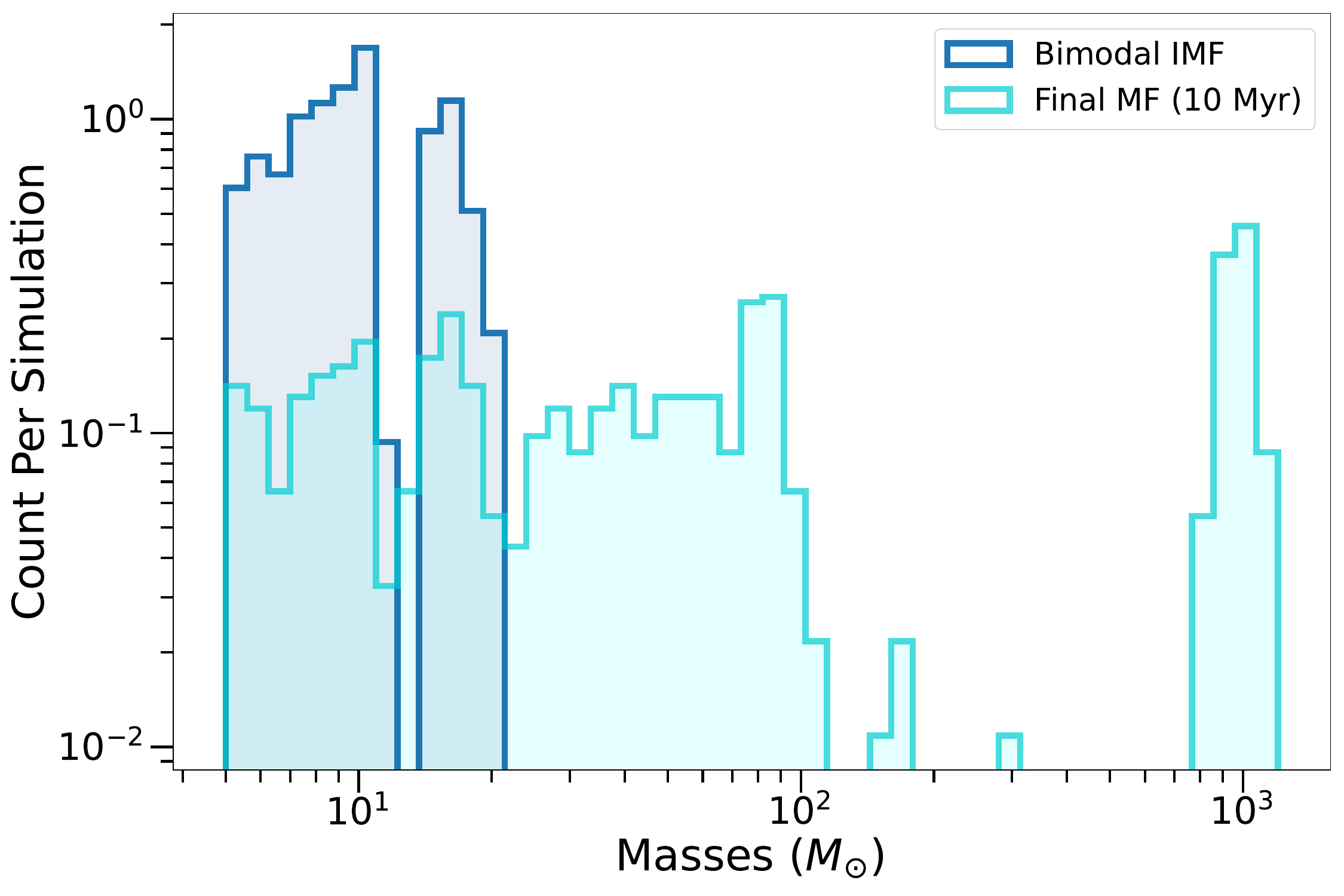}
    \caption{Comparison of the initial {\em (dark blue)} and final {\em (light blue)} MFs after 10 Myr averaged over the 90 bimodal simulations.}
    \label{fig:bimodalMF}
\end{figure}

The Salpeter IMF used in the simulations described in the previous section has the advantage of being directly empirically fit to observed stellar populations. However, the BH population is likely fundamentally different from stellar populations. Therefore, we also perform a set of 90 N-body simulations using an alternative BH IMF based on stellar evolution modeled with {\sc mesa} by \cite{Schneider_2023}. We call this IMF our bimodal IMF because, as Figure \ref{fig:3IMFs} shows, it has two peaks at 9 and 16~$M_\odot$ (see \S~\ref{sec:imf_choices} for more details).

Figure \ref{fig:bimodalMF} shows the initial and final MFs for our simulations with this bimodal IMF. Despite how different the bimodal IMF is from the Salpeter IMF, above $\sim 30$~$M_{\odot}$ the final MFs for the two IMFs are qualitatively similar. The main differences in this mass range are a smaller peak around $35~M_\odot$ and a larger peak around $80$~$M_{\odot}$. Larger differences in the final MFs of our different IMF simulations occur at $<30$~$M_{\odot}$. Although the initial peaks in the bimodal IMF are far less significant after 10~Myr, they are still clearly present at 9 and $16~M_\odot$. The Salpeter IMF, on the other hand, has peaks at 5--8~$M_\odot$ and 11--15~$M_\odot$, resulting from the negative slope of the IMF and our low mass cutoff at $5~M_\odot$. The initial chirp mass dip in the bimodal IMF at $12$~$M_{\odot}$ remains only partially filled by BBH mergers, because merger products with masses $<12$~$M_{\odot}$ are even rarer for bimodal simulations due to the positive low mass slope of the IMF.

We show the average distribution of primary masses for the bimodal IMF simulations in light blue in Figure \ref{fig:primary_log}. As in the final MF we see peaks around $m_1=9~M_\odot$ and $m_1=16$--$18~M_\odot$, as well as a smaller peak around $m_1=25~M_\odot$, which results from the high initial abundance of 9 and $16~M_\odot$ BHs producing a higher abundance of first generation merger products around $18$ and $25~M_\odot$. The top right panel of Figure \ref{fig:qvMass_histogram} shows a 2D histogram of the distribution of $q$ and $M_{\rm bin}$ for all BBH mergers over 10~Myr for the 90 bimodal IMF simulations. There are more peaks and valleys in this distribution than in the Salpeter simulations due to the IMF peaks at 9 and $16~M_\odot$. However, more broadly, the distribution of $q$ and $M_{\rm bin}$ for both IMF simulations are similar and have the same four over-densities (see \S~\ref{sec:mq}).

\subsection{Testing Merger Criteria}
\label{sec:criteria}

\begin{figure*}
\includegraphics[width=\columnwidth]{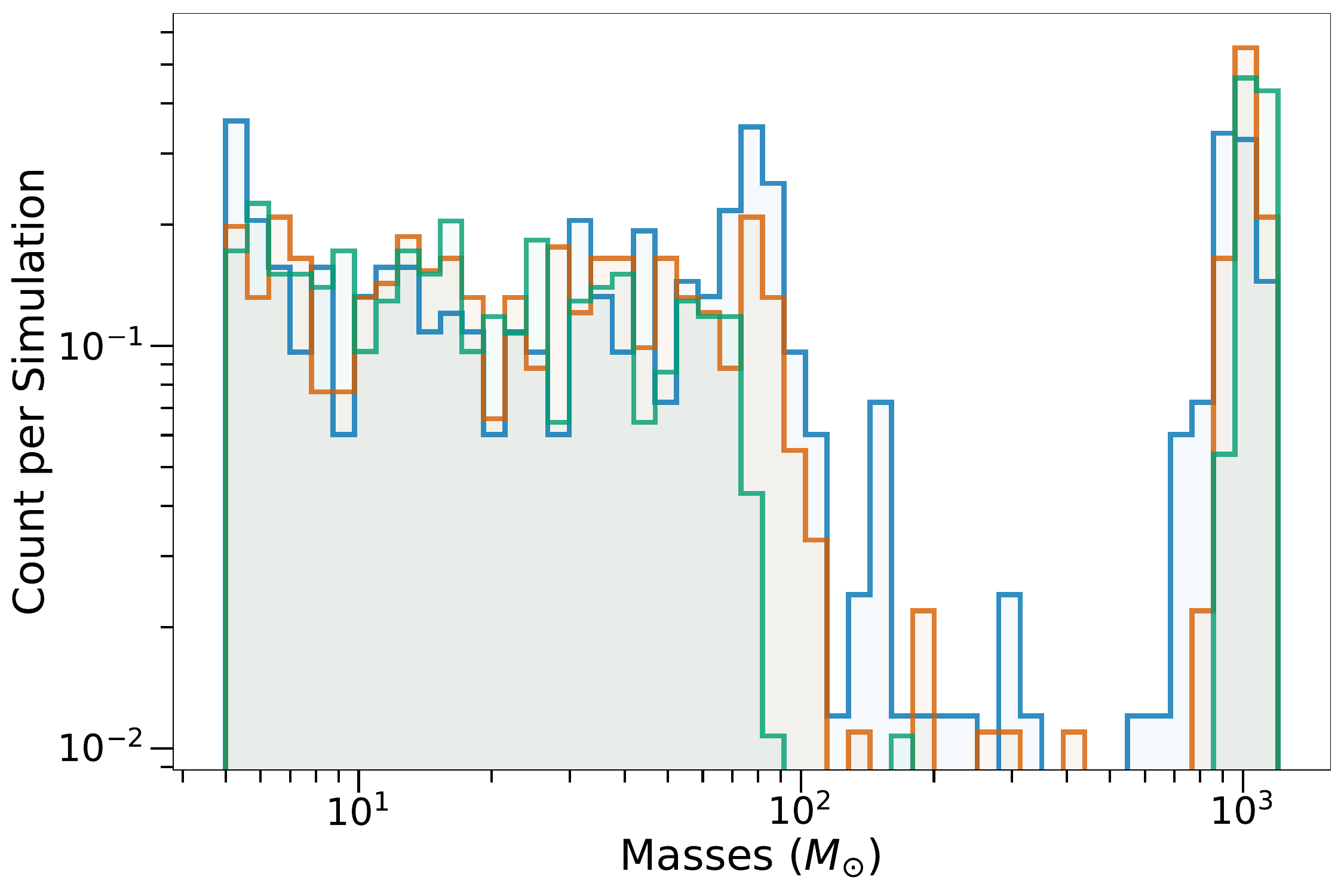}
\includegraphics[width=\columnwidth]{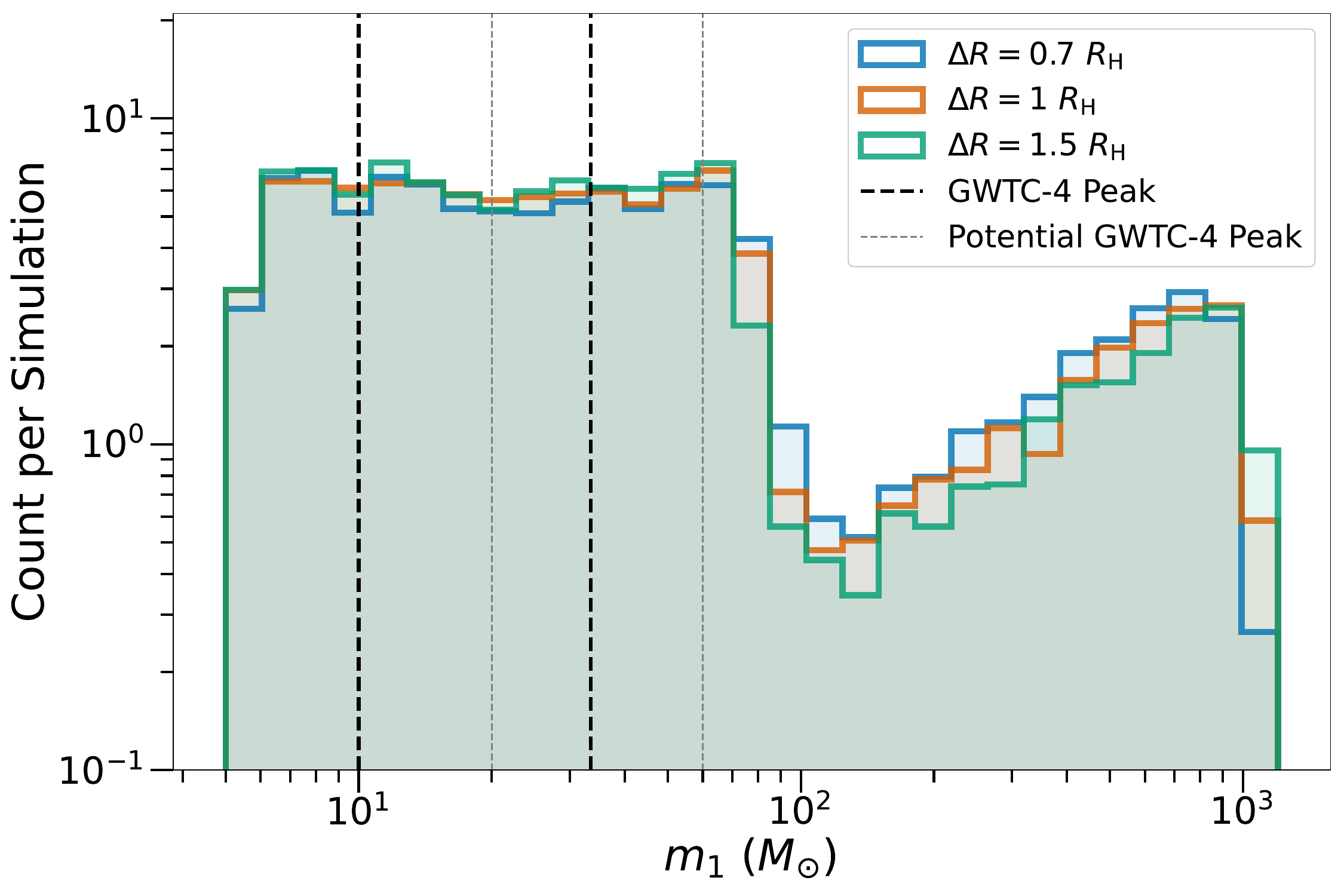}
    \caption{Comparing the final MFs after 10~Myr {\em (left)} and the primary MFs of BBH mergers {\em (right)} for Salpeter simulations where we change our merger criterion for the separation between merger components from the standard $\Delta R= 1\,R_{\rm H}$ (in orange) to $\Delta R= 0.7\,R_{\rm H}$ (in blue) or $\Delta R= 1.5\,R_{\rm H}$ (in green). For comparison, dashed lines show the location of peaks reported in GWTC-4.}
    \label{fig:Rhill_All}
\end{figure*}

Our simulated MFs depend on our prescriptions for BBH formation and mergers. In \citetalias{Secunda_2019} we estimated that the time for a gas-assisted BBH merger in an AGN disk is significantly less than the migration timescale, suggesting that multi-body interactions are unlikely to prevent a BBH from merging. Therefore, we assume in this paper that a BBH will merge within a negligibly short timescale if two BHs are within one mutual Hill sphere ($\Delta R= 1\,R_{\rm H}$) and have a relative kinetic energy lower than their binding energy (see \S~\ref{sec:nbody}). In this section, we conduct two additional sets of simulations where we change the distance required between two BHs for a merger to better understand the sensitivity of BH MFs to different merger criteria. In one set of simulations we test making our merger criteria stricter by requiring the distance between the two BHs to be $\Delta R= 0.7\,R_{\rm H}$. In the second set of simulations we allow two BHs to merge at $\Delta R= 1.5\,R_{\rm H}$, because recent hydrodynamic simulations suggest that wider separations are more favorable for BBH formation in AGN disks, \citep[e.g.,][]{Whitehead25}.
 
We compare the final MFs and primary MFs of simulations with our different merger criteria in the left and right panels of Figure~\ref{fig:Rhill_All}, respectively. The merger histories, final MFs, and primary MFs of our simulations with different merger criteria are qualitatively consistent with each other. Regardless of the required BH separation for a merger to occur, low mass BHs are depleted, the upper mass gap is filled, and IMBHs form. In addition, most of the BBH mergers have primary masses $5~M_\odot<m_1<70~M_\odot$ and there are also IMBH mergers with typical primary masses $400~M_\odot<m_1<1000~M_\odot$. 

However, there are several smaller differences that occur when we change the merger criteria. For example, our simulations with the smaller $\Delta R= 0.7\,R_{\rm H}$ criterion form fewer, less massive IMBHs and have a less distinct IMBH mass gap at $80~M_\odot<m_1<800~M_\odot$. These differences occur because it is more difficult for low mass objects to merge quickly, slowing the evolution of the MF of BHs in our AGN disk. As a result, the final MF for these simulations resembles the MF from an earlier epoch in our simulations with our standard merger criteria. On the other hand, in simulations with the larger $\Delta R= 1.5\,R_{\rm H}$ criterion, BHs merge more quickly, forming more massive IMBHs and having a more distinct mass gap. These distinctions emerge even though the relative kinetic energy condition may be more difficult to fulfill at larger BH separation.

The bottom panels of Figure \ref{fig:qvMass_histogram} show that the distributions of binary masses and mass ratios remain mostly consistent as well when we change our merger criteria. The largest differences come from the longer survival of less massive BHs in simulations with the $\Delta R= 0.7\,R_{\rm H}$ merger criterion, leading to a smaller dearth of mergers with $q<0.5$ around $100$--$300~M_\odot$ and slightly more mergers with $M_{\rm bin}>200$ and $q<0.1$ for these simulations. Conversely, the fast consumption of lower mass BHs in simulations with the $\Delta R= 1.5\,R_{\rm H}$ criterion produces a larger dearth of mergers with $q<0.5$ around $100$--$300~M_\odot$ and fewer mergers with with $M_{\rm bin}>200$ and $q<0.1$.

\section{Discussion}

\subsection{Comparison to GWTC-4}
\subsubsection{Mass Functions}
\label{sec:compare_gwtc3}

\begin{figure}
\includegraphics[width=\columnwidth]{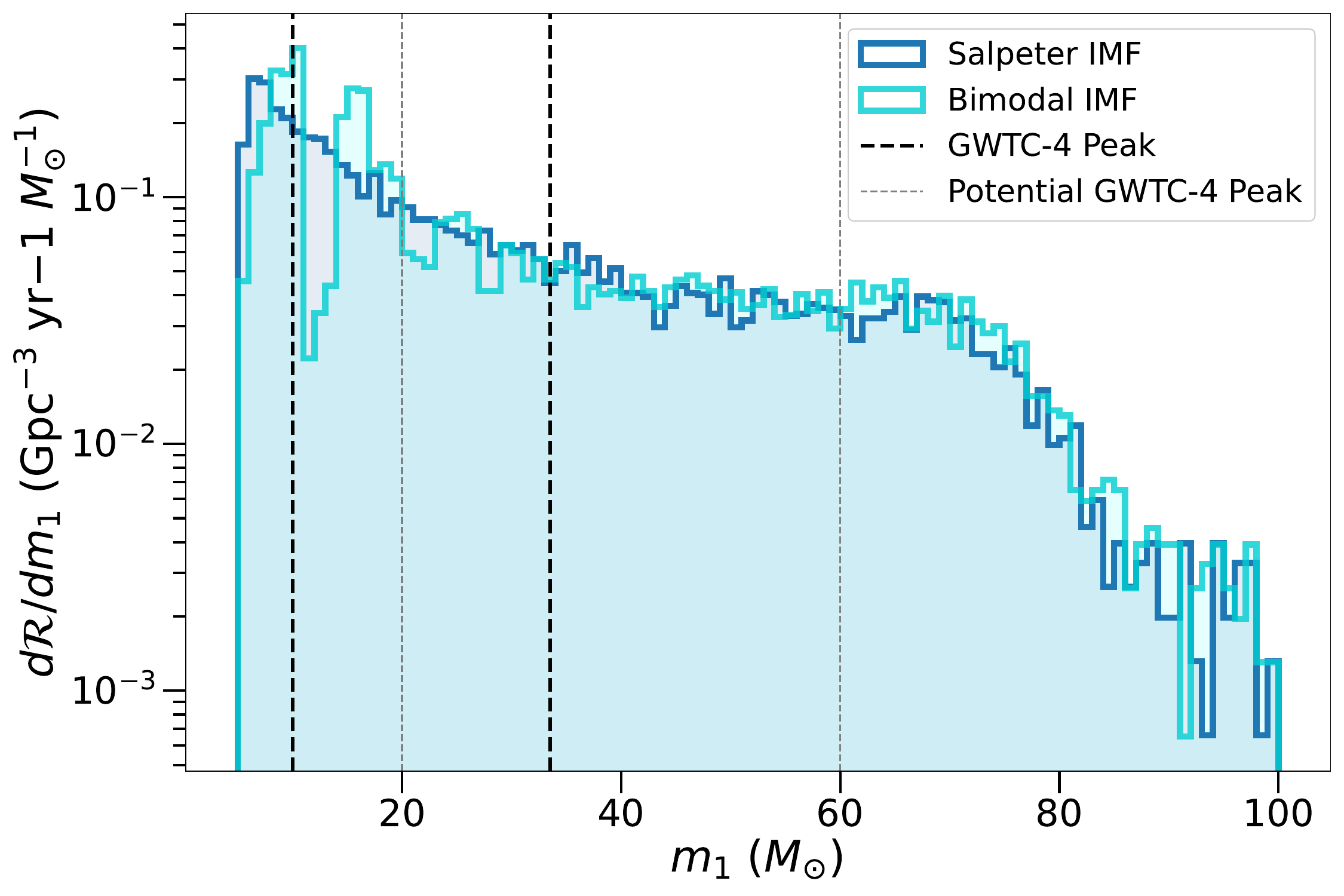}
    \caption{The merger rate as a function of primary mass over 10~Myr for simulations initialized with a Salpeter {\em (dark blue)} and bimodal {\em (light blue)} IMF. We zoom into the primary mass range shown in Figure 3 of \citet{GWTC4} and give the volumetric rate using the parameterization in Equation (\ref{eq:rates}). For comparison, dashed lines show the location of peaks reported in GWTC-4.}
    \label{fig:MF_GWTC4}
\end{figure}

In this section, we compare the primary MFs from our two different IMF simulations to the GWTC-4 primary MF \citep{GWTC4}. Although this mass spectrum is no doubt the convolution of multiple BBH merger channels, our goal here is to compare apparent features in our MFs with LVK observations to understand how relevant the AGN contribution might be. We show the merger rate as a function of primary mass over the mass range probed by GWTC-4 in Figure \ref{fig:MF_GWTC4}. Following \cite{Mckernan_2018}, we parameterize the rate of BBH mergers in an AGN disk as
\begin{equation}
\label{eq:rates}
\begin{split}
R = 12 \mbox{ Gpc}^{-3}\mbox{
  yr}^{-1}\frac{N_{\rm GN}}{0.006\mbox{ Mpc}^{-3}}\frac{N_{\rm BH}}{2\times10^4}\frac{f_{\rm AGN}}{0.1} \\
   \times \frac{f_{\rm d}}{0.1}\frac{f_{\rm b}}{0.1}\frac{\epsilon}{1}\left(\frac{\tau_{\rm AGN}}{10
  \mbox{ Myr}}\right)^{-1},
\end{split}
\end{equation}
where $N_{\rm GN}$ is the average number density of galactic nuclei in the Universe, $N_{\rm BH}$ is the number of BHs in a nuclear cluster, $f_{\rm AGN}$ is the fraction of galactic nuclei with AGN that last for time $\tau_{\rm AGN}$, $f_{\rm d}$ is the fraction of BHs that end up in the AGN disk, $f_{\rm b}$ is the fraction of BHs that form binaries, and $\epsilon$ represents the fractional change in $N_{\rm BH}$ over one full AGN duty cycle. To estimate the rate in each mass bin, we substitute in the number of BBHs in each bin for $N_{\rm BH}f_{\rm d}f_{\rm b}$.

Rapid hierarchical mergers in our simulations using both IMFs fill the mass gap above 40~$M_{\odot}$ in $<0.5$~Myr. These hierarchical mergers lead to a shallower slope above $35~M_\odot$ than in the GWTC-4 MF. As a result, our simulations predict a merger rate that is lower than the GWTC-4 rate for all primary masses $m_1<45~M_\odot$, but is comparable to the GWTC-4 rate at $m_1>45~M_\odot$. 

The GWTC-4 primary MF has two prominent peaks at $\sim 10~M_{\odot}$ and $\sim 33~M_{\odot}$ (shown as the darker dashed lines in Figure \ref{fig:MF_GWTC4}) and there is evidence of additional peaks at $\sim 20~M_{\odot}$ and $\sim 60~M_{\odot}$ (shown as the lighter dashed lines in Figure \ref{fig:MF_GWTC4}). The only GWTC-4 peak we observe in the primary MF for both of our IMF simulations is this last peak at $60$--$70~M_{\odot}$. This peak is a direct consequence of the threshold mass we find for perturbing an orbiter in a resonant orbit around the BH in the migration trap out of resonance. Therefore, this peak is entirely unrelated to the IMF, making it a robust prediction of migration-trap-assisted BBH mergers.

In the Salpeter simulations the primary MF peaks short of the observed $10~M_\odot$ peak, and there is no significant peak at either $m_1=20~M_\odot$ or $35~M_{\odot}$. Our simulations using the bimodal IMF produce a primary MF that is more consistent with GWTC-4 with peaks at $m_1=10~M_{\odot}$ and $15<m_1<19~M_{\odot}$, as in the GWTC-4 MF. These peaks in the bimodal primary MF are a direct result of the bimodal IMF instead of a result of the AGN environment, but it is relevant that the AGN environment does not completely destroy these IMF features.

Overall, the primary MFs from our simulations replicate some, but not all, of the specific features in the GWTC-4 primary mass spectrum. The bimodal IMF appears to do a better job of capturing the low mass spectrum, but neither IMF has a significant peak at $35~M_\odot$. The hierarchical merger fraction of the LVK peak at $35~M_{\odot}$ could be tested by searching for BHs in the peak with high spin \citep{Roy25}, which in turn would allow us to constrain the IMF producing such masses. For example, this peak would be more prominent in an AGN disk if the IMF in AGN disks was highly mass segregated \citep{RomSari25,Delfavero25b}. Heavier first generation mergers producing a peak at $35~M_\odot$ would also decrease the production of BHs in the range $\sim 45$--60~$M_{\odot}$, and yield identifiable ``echo merger" peaks at $\sim 70$, 105, and $140~M_{\odot}$ \citep{McKernan:2025,FM25}. Better statistics from O5 and other future LVK observing simulations on the possible peak at $\sim 70~M_{\odot}$ in the BH mass spectrum could both help distinguish between different IMFs and test the importance of migration trap mergers in AGNs.

Given our choice of IMFs, our mass spectrum would also be altered if the rate of orbiters entering the inner 1000~$R_g$ of the AGN disk via either disk capture of initially inclined orbiters from the nuclear star cluster \citep{Fabj20,Nasim23,Generozov23} or migration from the outer disk is mass-dependent or higher than proposed here. In our current simulations, we fix the rate of incoming orbiters to be constant as a function of mass, but in fact the rate of both disk capture and migration should increase as a function of mass \citep{Goldreich1979,Ward1997,Nasim23,Rowan:2025}. Adding mass dependence could lead to larger peaks at $10$, $20$ and $35~M_\odot$.

Alternatively, our merger channel may not play an important role in the LVK MF at masses $<40$~$M_\odot$. There is some indication in GWTC-4 that the BBH mergers that make up the peak at $33~M_\odot$ tend to have lower spins \citep{GWTC4}, which is consistent with our results here that this peak is not related to BBHs merging in AGN disks. Instead, the LVK MF $<40$~$M_\odot$ may be accounted for through isolated binary formation channels that can accurately predict the structure in the LVK MF at masses $<40$~$M_\odot$ \citep[e.g.][]{VanSon_2022,VanSon_2023}. On the other hand, our simulated AGN disk is highly efficient at forming over-massive BHs in the upper mass gap, which are present in LVK detections but hard to produce in isolated binary channels. For example, our simulations easily produce the 58 and $95~M_\odot$ components of GW231028\_153006 \citep{GWTC4a}.

\subsubsection{Rates}


In addition to comparing the properties of the primary mass functions in our N-body simulations to the GWTC-4 primary mass function, we also compare the rates of BBH mergers in our 360 simulations to the predicted GWTC-4 rates using Equation \ref{eq:rates}. The average number of BBH mergers in our simulations is nearly consistent between our simulations with Salpeter and bimodal IMFs, $103\pm2.3$ (for the Salpeter simulations). This value gives an overall rate of migration-trap-assisted BBH mergers of $6.20\pm 0.14$~Gpc$^{-3}$~yr$^{-1}$. The BBH merger rate predicted from GWTC-4 is 14--26~Gpc$^{-3}$~yr$^{-1}$ \citep{GWTC4}. Comparing these rates, we see that migration-trap-assisted mergers only account for a portion ($\sim 1/4$--1/2) of mergers detected by LVK, which is consistent with our conclusion in \S~\ref{sec:compare_gwtc3} above.

Roughly 40\% of the BBH mergers in our simulations are uneven mass ratio mergers with $q<0.5$ and $m_1<200$, giving a rate of $2.40\pm 0.39$~Gpc$^{-3}$~yr$^{-1}$. This rate is roughly consistent with the high end of the predicted rate of uneven mass ratio mergers in GWTC-4, suggesting that most of the uneven mass ratio mergers observed by LVK could come from an AGN disk. There is also evidence that $q$ may be anti-correlated with high effective spin parameters, $\chi_{\rm eff}$ \citep{Callister21,Adamcewicz23,Tagawa26}, which could be roughly consistent with predictions made for the distribution of BH spins in AGN disks (\citetalias{Secunda_2020}; \citealt{Mckernan_2018}; \citealt{Cook:2025}; \citealt{Su_2025}). 

In addition, migration-trap-assisted mergers could also account for many of the higher mass mergers detected by LVK. We find a rate of $1.65\pm0.25$~Gpc$^{-3}$~yr$^{-1}$ for mergers where $m_1 \in [20,50]\,M_{\odot}$ in our Salpeter simulations, which is a significant fraction of the corresponding rate predicted from GWTC-3, 2.5--6.3~Gpc$^{-3}$~yr$^{-1}$ \citep{Abbott:2023}. Our rate for the highest mass range of BHs in GWTC-3, $m_1 \in [50,100]\,M_{\odot}$, is $0.966\pm0.20$~Gpc$^{-3}$~yr$^{-1}$ for our Salpeter simulations, which is higher than the predicted rate from GWTC-3 of 0.099--0.4~Gpc$^{-3}$~yr$^{-1}$. These rates are consistent for our bimodal simulations. 

Both IMF simulations appear to overproduce high mass mergers, perhaps because of our assumption that all BBHs formed in our simulations will merge or because we have overestimated the parameter values in Equation (\ref{eq:rates}). Nonetheless, given our assumptions, an AGN disk could easily account for all of the BBH mergers with $m_1>50~M_\odot$. There is some evidence that higher mass BBHs with $m_1\gtrsim40~M_\odot$ have higher spins \citep[e.g.,][]{Wang:2022,Mould:2022,Godfrey:2023,Antonini:2025,Guo:2024,Li:2024,Li:2025,Sadiq:2025}, which would suggest these BBH mergers come from hierarchical formation channels. Further modeling of the spins among the high mass population in GWTC-4 could provide an upper limit on the AGN fraction of high mass events.


\subsection{Resonant Orbiters}
\label{sec:resonance}

\begin{figure}
    \centering
    \includegraphics[width=\columnwidth]{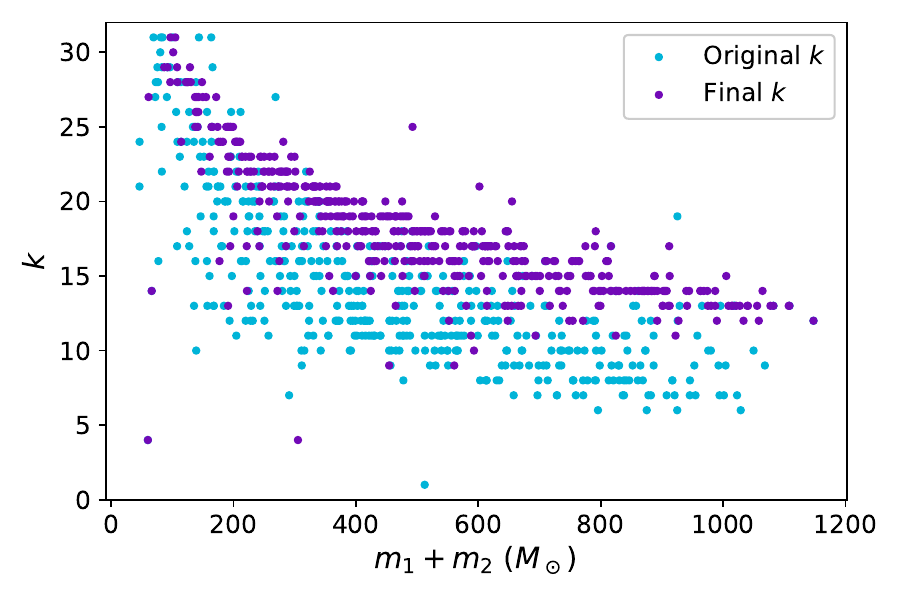}
    \caption{The harmonic $k$ of orbiters in first order resonances with the BH in the migration trap as a function of the total mass of the two orbiters, $m_1+m_2$, for all Salpeter simulations. Points in blue show the resonance harmonic of the initial resonant orbit, while the purple points show the last resonant harmonic before the resonance is broken due to resonant overlap leading to merger. Higher $k$ and total masses lead to resonant overlap.}
    \label{fig:resonance}
\end{figure}

Resonant in-plane orbits are a unique testable prediction of embedded objects in AGN disks, and may be detected in gravitational waves. In this section, we provide additional detail on the evolution of the resonant orbiters in our simulations. In \citetalias{Secunda_2019}, the majority of our simulations ended after 1~Myr with $1-3$ BHs just inside or outside the migration trap on resonant orbits with a $m\gtrsim50~M_\odot$ BH in the migration trap. These BHs were in first order resonances, $k+1:k$, with high harmonics $k$, such as the $k=27$ resonance shown in Figure 12 in \citetalias{Secunda_2019} (see also Figure 11 in \citetalias{Secunda_2019}).

The existence of 1--3 resonant orbiters at any given epoch persisted in \citetalias{Secunda_2020}, but adding orbiters migrating inward from the outer disk led to a different evolution for orbiters with semi-major axes larger or smaller than the migration trap radius. Orbiters at radii with semi-major axes smaller than the trap radius often remain in stable resonant orbits for millions of years. These orbiters only become destabilized due to stochastic fluctuations or scattering effects from multi-body interactions of orbiters at larger radii. For example, the yellow orbiter in Figure \ref{fig:singlerun} remains on a stable internal resonant orbit for 3~Myr before being disrupted by scattering. On the other hand, BHs in resonant orbits at semi-major axes larger than the trap radius typically merge with BHs migrating inward from the outer disk until they are perturbed out of resonance. Often there are two resonant orbiters in a chain outside the migration trap that can both merge with additional BHs migrating inward. These resonant orbiters typically merge with each other before merging with the BH in the migration trap.


In this paper, we again find that resonant orbiters play a significant role in our simulations, and we use the results from our 360 simulations to examine these resonances in more detail. Figure \ref{fig:resonance} shows the original and final harmonics for all orbiters in our Salpeter simulations in a first order resonance with the BH in the migration trap as a function of the total mass of the two orbiters. Repeated mergers with incoming orbiters push the resonant orbiters inward closer to the migration trap, leading to an increase in the resonant harmonic. Sometimes this inward motion brings the resonant orbiter to within a Hill radius of the BH in the migration trap, leading to a merger. Other times, before the BH passes within a Hill radius, the orbital resonance reaches a harmonic high enough to cause resonant overlap \citep{Wisdom1980}. This resonant overlap will break the resonance, also leading to a merger between the two BHs. We see in Figure \ref{fig:resonance} that as the mass of the BH in the migration trap grows, resonant overlap occurs at a lower harmonic, as predicted analytically \citep[e.g.,][]{Deck:2013}.

We find a constant threshold mass of $\sim80~M_\odot$ at which most of the resonant orbiters in our simulation merge with the BH in the migration trap. This mass is constant even though the BH in the migration trap grows from $100$--$1000~M_\odot$ over the course of the simulation, because as the BH in the migration trap grows, the initial width of the resonant orbits grows, leading to lower initial harmonics. This threshold mass produces a peak in the primary MF for BBH mergers in our simulations $\sim70~M_\odot$, which is roughly the final mass of the resonant orbiter for its last merger with an incoming BH. The threshold mass also leads to a steep drop in mergers with $70~M_\odot<m_1<300~M_\odot$, which we call the IMBH mass gap. We find that the threshold mass is important as long as enough additional orbiters are injected over the course of the simulation for resonant orbiters to reach this mass.

Our results here are consistent with the analytic criteria for the stability of mean motion resonances of BHs in AGN disks derived in \citet{Epstein-Martin2025} using disk models and torque prescriptions similar to those evolved in our N-body simulation. \citet{Epstein-Martin2025} find that mean motion resonances are only possible in the migration trap region for $M_{\rm SMBH}=10^8~M_\odot$ SMBHs for large BH masses and high harmonics, due to general-relativistic apsidal precession. Although we do not include general-relativistic effects in our simulation, the resonant orbiters in our simulations meet the criteria in \citet{Epstein-Martin2025}, because resonances only occur when the BH in the migration trap is $\gtrsim 50~M_\odot$ and we only find resonances with $k>10$ until the total mass of the orbiters is $>300~M_\odot$. We find our simulations favor high harmonics at small radii as \citet{Epstein-Martin2025} suggest.

\citet{Epstein-Martin2025} also derive the impact of scattering effects from the surrounding nuclear star cluster and stochastic turbulent fluctuations on resonant orbiters. We do not include scattering effects from inclined orbiters in our simulations, but the large disk volume due to the large SMBH mass should make scattering events rare \citep{Secunda:2021,Epstein-Martin2025}. We do find on occasion that the turbulence prescription we use here, which is different from \citet{Epstein-Martin2025}, breaks mean motion resonances. However, stochastic fluctuations are not the primary reason for resonance breaking in our simulations.

Increasing the level of turbulence would lead to more frequent resonance breaking. Fewer resonant orbiters, due to more frequent stochastic events than modeled here, could erase the peak in the primary MF at $m_1\approx70~M_\odot$. There would also be fewer $q= 1$--0.1 mergers with $M_{\rm bin}\approx100$--$1000~M_\odot$. These mergers would be replaced by more $q\approx 0.1$--0.01 mergers with $M_{\rm bin}\approx100$--1000, which do occur in our current simulation but are rarer. These BBH mergers could perhaps be observed as intermediate mass ratio inspirals (IMRIs) by LISA (see \S~\ref{sec:imbh}), leading to a higher IMRI rate for more frequent stochastic events.

Resonant orbiters may also produce extreme mass ratio inspirals (EMRIs) that could be observed by LISA. In particular, the migration trap will be closer to the SMBH in lower mass SMBHs \citep{Grishin24}. Such systems would correspond to gravitational wave frequencies, $\nu_{\rm GW} \sim \mathcal{O}(10^{-6})~{\rm Hz}~(M_{\rm SMBH}/10^{6}M_{\odot})^{1/2}~(R/10^{3}r_{g})^{-3/2}$, which could lie in the LISA band for modest SMBH mass and close-in migration traps ($<10^{3}r_{g}$). A chain of EMRIs on correlated orbits (offset gravitational wave frequencies) appearing in the same LISA volume because they originate from the same AGN would represent a remarkably strong constraint on internal AGN disk properties.


\subsection{Intermediate Mass Black Hole Formation}
\label{sec:imbh}
There are multiple hypotheses for how IMBHs form, including runaway collisions \citep{Zwart2002}, accretion \citep{Miller2002}, or primordial black holes \citep{Carr2021}. \citetalias{Secunda_2020} showed that IMBHs can also be formed through hierarchical mergers in the migration trap of AGN disks. Here, we find that our simulations consistently form IMBHs in the $900-1100~M_\odot$ range, regardless of IMF or merger criteria, extending the results in \citetalias{Secunda_2020}.

For all types of simulations, this massive IMBH population begins to form and break away from the distribution of masses less than $100~M_\odot$ at $\sim$2~Myr, creating an IMBH mass gap. The most notably different simulations are the ones in which we tighten our merger criteria by requiring the BH separation $\Delta R < 0.7$~$R_{\rm H}$ for two BHs to merge. For these simulations, the IMBH peak is slightly lower and the IMBH mass gap is less distinctive. Nonetheless, the overall consistency between the different types of simulations suggests that the formation of IMBHs is robust to these variations, and IMBHs with masses greater than 200~$M_\odot$ should form as long as the AGN disk lifetime is longer than 2~Myr.

The merger histories of our simulations, such as the example merger history in Figure \ref{fig:singlerun}, show that these IMBHs end up embedded in the migration trap. Once in the migration trap, they primarily grow from mergers with resonant orbiters that have built up to roughly 80~$M_{\odot}$ through mergers with other lower mass BHs. Therefore, we find that it is the existence of the migration trap specifically that supports the growth of IMBHs in AGN disks. Figure~\ref{fig:qvMass_histogram} shows that lower mass IMBHs may grow through relatively even mass mergers, but as the IMBH mass increases, the mergers become more uneven, $q<1$, because of the threshold mass of around 80~$M_{\odot}$ we find for resonant orbiters.

LVK is not capable of directly observing gravitational wave events involving BHs this massive. However, future instruments, such as LIGO A+, LISA, the Cosmic Explorer, and the Einstein Telescope, may be able to observe these mergers between IMBHs and stellar mass BHs, the light class of IMRIs \citep{LISA23}. Using Equation (\ref{eq:rates}), we find a light IMRI rate of $1.01\pm0.19$~Gpc$^{-3}$~yr$^{-1}$ in our simulations. 

The IMBHs in our simulations may become massive enough to open a gap in the gas disk and undergo rapid Type II migration \citep{Lin:1986}, potentially forming a heavy IMRI. An IMBH that forms close to the SMBH could also decay into the SMBH during quiescence (post AGN), producing a heavy IMRI \citep{LISA23}. One IMBH-SMBH inspiral per AGN disk lifetime gives a rate of 0.06~Gpc$^{-3}$~yr$^{-1}$ using Equation (\ref{eq:rates}). The rate of light IMRIs will be higher for a longer disk lifetime, while the rate of heavy IMRIs will be higher for a shorter disk lifetime, or if Type II migration is important in AGN disks. As a result, the rate of detection of IMBHs via light or heavy IMRIs with LISA and future gravitational wave detectors can be used as a direct probe of the average lifetime of dense AGN disks in our Universe, where in AGN disks IMBHs can form, the existence of migration traps, and the importance of Type II migration in AGN disks.

\subsection{Comparison to Monte Carlo Simulations}
\label{sec:mc}
Most recent studies of the evolution of more than three orbiters embedded in an AGN disk use semi-analytic Monte Carlo methods. Monte Carlo methods are significantly less computationally expensive than N-body simulations, allowing for a broader parameter survey. However, the advantage of our augmented N-body simulations is that they explicitly track the phase-space evolution of orbiters under the influence of gravitational and gas forces. Because the two types of simulation are complementary to each other, we will now compare our results to results from recent Monte Carlo simulations.

We will first compare our results with the results from simulations using the \texttt{McFACTS} Monte Carlo code \citep{McKernan:2025}. \texttt{McFACTS} uses a default IMF similar to the Salpeter IMF used here and a similar default disk model. However, \texttt{McFACTS} simulates the AGN disk out to $>10^4~R_g$ and uses a default disk lifetime of only $\sim 0.5$~Myr, because the maximal merger rate for the AGN channel comes from  multiple short-lived AGN episodes in the same plane \citep{FM25}. Due to the short lifetime of their disk, the \texttt{McFACTS} final MF for one AGN episode most closely resembles the MF found here after 0.5~Myr, although there are several differences. The largest difference is that the final MF for the default \texttt{McFACTS} IMF has a much steeper drop off in mass, especially above $35~M_\odot$, where instead the 0.5~Myr MF from our Salpeter simulations has a peak from $30-80~M_\odot$. This difference is a direct result of \texttt{McFACTS} not explicitly integrating the gravitational evolution, preventing resonant orbiters from building up the way we see in our N-body simulations. 

The \texttt{McFACTS} simulations also produce IMBHs. However, \cite{McKernan:2025} find that more IMBHs are produced in the migration swamp that forms around $10^3~R_g$ than in the migration trap for their default disk model. This migration swamp is a region where the migration rate slows without changing signs, leading to a pile up of orbiters. We find, using a similar disk model and migration rates, that the region closest to the migration trap is still dominant in the process of producing IMBHs. However, if we expand the outer radius of our simulations to $>10^4~R_g$ we might also observe a similar traffic jam. \cite{McKernan:2025} find that together migration traps and swamps account for $\sim 20\%$ of all mergers, or a rate of 4~Gpc$^{-3}$~yr$^{-1}$ for migration swamp-trap mergers. This rate is somewhat lower than the one we find for our simulations of the same region at $<10^3~R_g$.

 \citet{Vaccaro25} test the fraction of events that occur at or near migration traps for SMBH masses in the range $M_{\rm SMBH}=10^{5}$--$10^{9}\,M_{\odot}$. Unlike in \citet{McKernan:2025}, they find that mergers in migration traps can account for up to $\sim 80\%$ of the total rate, for SMBH masses up to $M_{\rm SMBH}\sim 10^{8}M_{\odot}$. However, \citet{Vaccaro25} find that migration traps can disappear from disk models above $M_{\rm SMBH} \sim 10^{8}M_{\odot}$, following \citet{Grishin24} and \citet{Gilbaum:2025}, and that a change in the viscosity prescription of the disk can change the location of migration traps. 
 
\subsection{Caveats}

In this section, we briefly outline several caveats to the results presented in this paper. One of the main assumptions in this paper is that our merger criteria are a decent prescription for the requirements for BBHs to form and merge in AGN disks. In \citetalias{Secunda_2019} we showed that the timescales for a gas-assisted merger for the BBHs formed in our simulation are shorter than the migration timescales, suggesting that multi-body interactions would not play a significant role in disrupting BBH mergers. In \S~\ref{sec:criteria} we now directly test the impact of our merger criteria on our results by using several different BBH formation criteria inspired by recent hydrodynamic simulations of BBH formation \citep{Rowan2023,Qian2024,Whitehead2024,DeLaurentiis2023,Whitehead25}. 

Figure \ref{fig:Rhill_All} shows that a wider merger criterion that increases the effective radius for BBH capture allows easier consumption of lower-mass BHs, leading to the formation of larger IMBHs and mildly accelerating hierarchical BH growth. On the other hand, a narrower merger criterion slows the formation of IMBHs, leading to a less prominent IMBH mass gap at $80~M_\odot<m_1<800~M_\odot$ and lower mass IMBHs, as occurs in earlier epochs of our simulations with standard merger criteria. Overall, tightening and relaxing the merger criterion leads to similar final MFs and primary MFs, which suggests that our qualitative results are robust and not due to a specific choice of capture radius. Future work should study the evolution of BBHs down to $\Delta R<0.7~R_{\rm H}$ to further test for convergence.

The lifetime of an AGN disk is also uncertain; it could range from $10^5$--$10^7$~yr \citep{King:2015,Schawinski:2015,Marconi:2004}. To account for this uncertainty, we examine the MF at several different epochs. By 0.5~Myr, hierarchical mergers in our simulation have filled the mass gap to $> 100~M_{\odot}$ and the number of low mass BHs has already been depleted through mergers. Still, the primary MF for BBH mergers in Figure \ref{fig:evolutionhist} is narrower after only 0.5~Myr than at later epochs, peaking around $20~M_\odot$ and dropping off around $35~M_\odot$. Therefore, if AGN disks have a lifetime $\lesssim0.5$~Myr, migration-trap-assisted mergers may not be able to account for the larger mass mergers observed by LIGO. However, if the AGN disk lifetime is short but episodic, then the higher final MF of each disk episode could lead to higher mass mergers in subsequent disk episodes.

The largest difference between the MFs at earlier and later times in our simulation is that the IMBH population does not break off from the main mass distribution, forming the IMBH mass gap, until around 2~Myr. In addition, the mean mass of this population grows significantly between 3 and 10~Myr. The difference between IMBH masses at different epochs may be observable with LISA and allow us to estimate the lifetimes of AGN disks.

We make several other simplifications in our simulations that need closer examination in future work. These simplifications include using a constant, mass-independent rate for the addition of BHs to our simulations due to migration and disk capture, neglecting the growth of BHs via the accretion of gas and dust, merger kicks for BBH merger products, the influence of thermal feedback from the BHs on migration rates \citep{Benitez-Llambay_2015}, and the influence of retrograde orbiters \citep{Secunda:2021}.
We model the gas forces analytically, but future work should replace our analytic models with a gas dynamical or magnetohydrodynamical model.

Finally, investigating a wider range of AGN parameters, such as accretion rates and SMBH masses, is needed to understand the impact they will have on BBH formation. \citet{Delfavero25} explored a wider range of parameters using the Monte Carlo code \texttt{McFACTS}, and found that more BBH mergers occur in AGN disks with SMBH masses around $10^{8}\, M_\odot$, which supports our choice of SMBH mass here. In addition, detailed migration torque calculations suggest that migration traps can disappear from disk models with $M_{\rm SMBH} \gtrsim 10^{8}M_{\odot}$ \citep{Grishin24,Gilbaum:2025}.

\section{Conclusions}


Most recent AGN merger channel studies use either Monte Carlo simulations or hydrodynamical simulations of one to three BHs. In this paper, we use 3-D N-body simulations that have been augmented to include gas forces on orbiters. While computationally expensive, the advantage of 3-D N-body simulations is that they explicitly track the phase-space evolution of the BHs under the influence of the gravitational and gas forces. We investigate how different BH IMFs evolve over 10~Myr through hierarchical mergers in the migration trap region of dense AGN disks. We accumulate large-number statistics by running 360 N-body simulations, each containing $\sim100$ orbiters. Using two different IMFs and experimenting with three different sets of merger criteria, we find the following results.
\begin{enumerate}
    \item Hierarchical BH mergers fill the upper mass gap up to roughly 100~$M_{\odot}$ in $<0.5$~Myr, consistent with Monte Carlo results \citep[e.g.][]{McKernan:2025}. The BBH merger rate exceeds the estimated combined rate of replenishment of BHs through disk capture of inclined orbiters and migration, leading to fewer low mass BHs in the final MFs than in the IMFs and a nearly flat mass distribution between 15 and $80~M_\odot$.
    \item The primary MF for our bimodal IMF simulations more closely resembles the GWTC-4 primary MF at low masses, with peaks at 10 and $20~M_\odot$. 
    \item Runs with both IMFs generate a peak in the primary MF at $m_1\approx 70 M_\odot$, due to the properties of the resonant orbiters formed in our simulations. This peak may be related to the potential peak at $\sim60$--$70~M_\odot$ observed in the GWTC-4 primary MF.
    \item We calculate a predicted merger rate for migration-trap-assisted mergers of $6.20\pm 0.14$~Gpc$^{-3}$~yr$^{-1}$ (or 1/4--1/2 the observed LVK rate). The most common BBH mergers in our simulations are equal mass, low mass mergers. 
    \item Roughly $40\%$ of the mergers in our simulations are uneven mass ratio mergers with $q<0.5$. In addition, 16\% of mergers in our simulations have a primary mass $\in[50-100]~M_\odot$, giving a merger rate of $0.966\pm0.20$~Gpc$^{-3}$~yr$^{-1}$. Therefore, AGN disks could easily be the source of BBH mergers observed by LVK that are difficult to produce through traditional stellar evolution channels.
    \item  AGN disks with a migration trap form a separate higher-mass IMBH population after $\sim 3$~Myr. This population grows in mass over time, forming IMBHs as massive as 1000~$M_\odot$. This IMBH mass spectrum should be observable with LISA.
    \item Allowing BBH formation at wider separations in our simulations, as suggested by hydrodynamical simulations of BBHs embedded in gas disks \citep{DeLaurentiis2023,Rowan2023,Qian2024,Dodici24}, mildly accelerates hierarchical BBH mergers, but does not lead to qualitatively different results.
\end{enumerate}

By the end of O5, LVK will observe over a thousand BBH mergers, allowing us to directly test models for the astrophysical origin of different BH populations from the distribution of $M_{\rm bin}$, $q$, and $\chi_{\rm eff}$. In addition, LISA and future generations of gravitational wave detectors will test whether IMBHs form in AGN disks at migration traps and test our predicted light IMRI rate of $1.01\pm 0.19$~Gpc$^{-3}$~yr$^{-1}$ and heavy IMRI rate of $0.06$~Gpc$^{-3}$~yr$^{-1}$. The ratio of these IMRI rates can help estimate the average lifetimes of AGN disks.

\begin{acknowledgements}

We thank the anonymous referee for a useful report that improved our analysis. The authors thank Dr. Tiger C. Lu for his advice on mean motion resonances. We thank Open Science Grid staff Rachel Lombard, Shomic Islaam, and Christina Koch for their diligent support of our simulation campaign. KLG thanks the Things in Disks, Yo! NYC (TIDYNYC) research group, AstroCom NYC, and the AMNH/CUNY REU program for providing an excellent and supportive research environment. The Center for Computational Astrophysics at the Flatiron Institute is supported by the Simons Foundation. M-MML was partly supported by NSF grant AST23-07950 and by grant NSF PHY-23-09135 to the Kavli Institute for Theoretical Physics (KITP). KESF \& BM are supported by NSF AST-2206096 and NSF AST-1831415 and Simons Foundation Grant 533845 as well as Simons Foundation sabbatical support. This work has received funding from the European Research Council (ERC) under the European Union’s Horizon 2020 research and innovation programme (Grant agreement No.\ 945806) and is supported by the Deutsche Forschungsgemeinschaft (DFG, German Research Foundation) under Germany’s Excellence Strategy EXC 2181/1-390900948 (the Heidelberg STRUCTURES Excellence Cluster).

\end{acknowledgements}




\bibliography{example} 

@ARTICLE{Epstein-Martin2025,
       author = {{Epstein-Martin}, Marguerite and {Stone}, Nicholas and {Becker}, Juliette},
        title = "{Mean Motion Resonances in AGN Disks}",
      journal = {arXiv e-prints},
         year = 2025,
        month = oct,
          eid = {arXiv:2510.12895},
        pages = {arXiv:2510.12895},
          doi = {10.48550/arXiv.2510.12895},
archivePrefix = {arXiv},
       eprint = {2510.12895},
 primaryClass = {astro-ph.HE},
       adsurl = {https://ui.adsabs.harvard.edu/abs/2025arXiv251012895E}
}

@ARTICLE{Wisdom1980,
       author = {{Wisdom}, J.},
        title = "{The resonance overlap criterion and the onset of stochastic behavior in the restricted three-body problem}",
      journal = {\aj},
         year = 1980,
        month = aug,
       volume = {85},
        pages = {1122-1133},
          doi = {10.1086/112778},
       adsurl = {https://ui.adsabs.harvard.edu/abs/1980AJ.....85.1122W}
}

@ARTICLE{Benitez-Llambay_2015,
       author = {{Ben{\'\i}tez-Llambay}, Pablo and {Masset}, Fr{\'e}d{\'e}ric and {Koenigsberger}, Gloria and {Szul{\'a}gyi}, Judit},
        title = "{Planet heating prevents inward migration of planetary cores}",
      journal = {\nat},
         year = 2015,
        month = apr,
       volume = {520},
       number = {7545},
        pages = {63-65},
          doi = {10.1038/nature14277},
archivePrefix = {arXiv},
       eprint = {1510.01778},
 primaryClass = {astro-ph.EP},
       adsurl = {https://ui.adsabs.harvard.edu/abs/2015Natur.520...63B}
}

@ARTICLE{WZL2024,
       author = {{Wang}, Yihan and {Zhu}, Zhaohuan and {Lin}, Douglas N.~C.},
        title = "{Stellar/BH population in AGN discs: direct binary formation from capture objects in nuclei clusters}",
      journal = {\mnras},
         year = 2024,
        month = mar,
       volume = {528},
       number = {3},
        pages = {4958-4975},
          doi = {10.1093/mnras/stae321},
archivePrefix = {arXiv},
       eprint = {2308.09129},
 primaryClass = {astro-ph.GA},
       adsurl = {https://ui.adsabs.harvard.edu/abs/2024MNRAS.528.4958W}
}

@ARTICLE{Su_2025,
       author = {{Su}, Yubo},
        title = "{A Possible Mass Ratio and Spin-Orbit Misalignment Correlation for Mergers of Binary Black Holes in Nuclear Star Clusters}",
      journal = {arXiv e-prints},
         year = 2025,
        month = jan,
          eid = {arXiv:2501.16258},
        pages = {arXiv:2501.16258},
          doi = {10.48550/arXiv.2501.16258},
archivePrefix = {arXiv},
       eprint = {2501.16258},
 primaryClass = {astro-ph.HE},
       adsurl = {https://ui.adsabs.harvard.edu/abs/2025arXiv250116258S}
}

@ARTICLE{Rowan:2025,
       author = {{Rowan}, Connar and {Whitehead}, Henry and {Kocsis}, Bence},
        title = "{Black hole merger rates in AGN: contribution from gas-captured binaries}",
      journal = {\mnras},
         year = 2025,
        month = nov,
          doi = {10.1093/mnras/staf1896},
archivePrefix = {arXiv},
       eprint = {2412.12086},
 primaryClass = {astro-ph.HE},
       adsurl = {https://ui.adsabs.harvard.edu/abs/2025MNRAS.tmp.1796R}
}

@ARTICLE{VanSon_2023,
       author = {{van Son}, L.~A.~C. and {de Mink}, S.~E. and {Chru{\'s}li{\'n}ska}, M. and {Conroy}, C. and {Pakmor}, R. and {Hernquist}, L.},
        title = "{The Locations of Features in the Mass Distribution of Merging Binary Black Holes Are Robust against Uncertainties in the Metallicity-dependent Cosmic Star Formation History}",
      journal = {\apj},
         year = 2023,
        month = may,
       volume = {948},
       number = {2},
          eid = {105},
        pages = {105},
          doi = {10.3847/1538-4357/acbf51},
archivePrefix = {arXiv},
       eprint = {2209.03385},
 primaryClass = {astro-ph.GA},
       adsurl = {https://ui.adsabs.harvard.edu/abs/2023ApJ...948..105V}
}

@ARTICLE{VanSon_2022,
       author = {{van Son}, L.~A.~C. and {de Mink}, S.~E. and {Renzo}, M. and {Justham}, S. and {Zapartas}, E. and {Breivik}, K. and {Callister}, T. and {Farr}, W.~M. and {Conroy}, C.},
        title = "{No Peaks without Valleys: The Stable Mass Transfer Channel for Gravitational-wave Sources in Light of the Neutron Star-Black Hole Mass Gap}",
      journal = {\apj},
         year = 2022,
        month = dec,
       volume = {940},
       number = {2},
          eid = {184},
        pages = {184},
          doi = {10.3847/1538-4357/ac9b0a},
archivePrefix = {arXiv},
       eprint = {2209.13609},
 primaryClass = {astro-ph.HE},
       adsurl = {https://ui.adsabs.harvard.edu/abs/2022ApJ...940..184V}
}

@ARTICLE{VanSon_2020,
       author = {{van Son}, L.~A.~C. and {De Mink}, S.~E. and {Broekgaarden}, F.~S. and {Renzo}, M. and {Justham}, S. and {Laplace}, E. and {Mor{\'a}n-Fraile}, J. and {Hendriks}, D.~D. and {Farmer}, R.},
        title = "{Polluting the Pair-instability Mass Gap for Binary Black Holes through Super-Eddington Accretion in Isolated Binaries}",
      journal = {\apj},
         year = 2020,
        month = jul,
       volume = {897},
       number = {1},
          eid = {100},
        pages = {100},
          doi = {10.3847/1538-4357/ab9809},
archivePrefix = {arXiv},
       eprint = {2004.05187},
 primaryClass = {astro-ph.HE},
       adsurl = {https://ui.adsabs.harvard.edu/abs/2020ApJ...897..100V}
}

@ARTICLE{Broekgaarden_2022,
       author = {{Broekgaarden}, Floor S. and {Berger}, Edo and {Stevenson}, Simon and {Justham}, Stephen and {Mandel}, Ilya and {Chru{\'s}li{\'n}ska}, Martyna and {van Son}, Lieke A.~C. and {Wagg}, Tom and {Vigna-G{\'o}mez}, Alejandro and {de Mink}, Selma E. and {Chattopadhyay}, Debatri and {Neijssel}, Coenraad J.},
        title = "{Impact of massive binary star and cosmic evolution on gravitational wave observations - II. Double compact object rates and properties}",
      journal = {\mnras},
         year = 2022,
        month = nov,
       volume = {516},
       number = {4},
        pages = {5737-5761},
          doi = {10.1093/mnras/stac1677},
archivePrefix = {arXiv},
       eprint = {2112.05763},
 primaryClass = {astro-ph.HE},
       adsurl = {https://ui.adsabs.harvard.edu/abs/2022MNRAS.516.5737B}
}

@ARTICLE{Renzo_2024,
       author = {{Renzo}, M. and {Smith}, N.},
        title = "{Pair-instability evolution and explosions in massive stars}",
      journal = {arXiv e-prints},
         year = 2024,
        month = jul,
          eid = {arXiv:2407.16113},
        pages = {arXiv:2407.16113},
          doi = {10.48550/arXiv.2407.16113},
archivePrefix = {arXiv},
       eprint = {2407.16113},
 primaryClass = {astro-ph.HE},
       adsurl = {https://ui.adsabs.harvard.edu/abs/2024arXiv240716113R}
}

@ARTICLE{Gilbaum:2025,
       author = {{Gilbaum}, Shmuel and {Grishin}, Evgeni and {Stone}, Nicholas C. and {Mandel}, Ilya},
        title = "{How to Escape from a Trap: Outcomes of Repeated Black Hole Mergers in Active Galactic Nuclei}",
      journal = {\apjl},
         year = 2025,
        month = mar,
       volume = {982},
       number = {1},
          eid = {L13},
        pages = {L13},
          doi = {10.3847/2041-8213/adb7dc},
archivePrefix = {arXiv},
       eprint = {2410.19904},
 primaryClass = {astro-ph.HE},
       adsurl = {https://ui.adsabs.harvard.edu/abs/2025ApJ...982L..13G}
}

@ARTICLE{McKernan:2025,
       author = {{McKernan}, Barry and {Ford}, K.~E. Saavik and {Cook}, Harrison E. and {Delfavero}, Vera and {McPike}, Emily and {Nathaniel}, Kaila and {Postiglione}, Jake and {Ray}, Shawn and {O'Shaughnessy}, Richard},
        title = "{McFACTS I: Testing the LVK AGN Channel with Monte Carlo for AGN Channel Testing and Simulation (McFACTS)}",
      journal = {\apj},
         year = 2025,
        month = sep,
       volume = {990},
       number = {2},
          eid = {217},
        pages = {217},
          doi = {10.3847/1538-4357/adf114},
archivePrefix = {arXiv},
       eprint = {2410.16515},
 primaryClass = {astro-ph.HE},
       adsurl = {https://ui.adsabs.harvard.edu/abs/2025ApJ...990..217M}
}

@ARTICLE{Cook:2025,
       author = {{Cook}, Harrison E. and {McKernan}, Barry and {Ford}, K.~E. Saavik and {Delfavero}, Vera and {Nathaniel}, Kaila and {Postiglione}, Jake and {Ray}, Shawn and {McPike}, Emily J. and {O'Shaughnessy}, Richard},
        title = "{McFACTS. II. Mass Ratio─Effective Spin Relationship of Black Hole Mergers in the Active Galactic Nucleus Channel}",
      journal = {\apj},
         year = 2025,
        month = nov,
       volume = {993},
       number = {2},
          eid = {163},
        pages = {163},
          doi = {10.3847/1538-4357/adfd56},
archivePrefix = {arXiv},
       eprint = {2411.10590},
 primaryClass = {astro-ph.HE},
       adsurl = {https://ui.adsabs.harvard.edu/abs/2025ApJ...993..163C}
}

@ARTICLE{GWTC4,
       author = {{Abac}, A.~G. and {Abouelfettouh}, I. and {Acernese}, F. and {Ackley}, K. and {Adamcewicz}, C. and {Adhicary}, S. and {Adhikari}, D. and {Adhikari}, N. and {Adhikari}, R.~X. and {Adkins}, V.~K. and {Afroz}, S. and {Agarwal}, D. and {Agathos}, M. and {Aghaei Abchouyeh}, M. and {Aguiar}, O.~D. and {Ahmadzadeh}, S. and {Aiello}, L. and {Ain}, A. and {Ajith}, P. and {Akutsu}, T. and {Albanesi}, S. and {Alfaidi}, R.~A. and {Al-Jodah}, A. and {All{\'e}n{\'e}}, C. and {Allocca}, A. and {Al-Shammari}, S. and {Altin}, P.~A. and {Alvarez-Lopez}, S. and {Amarasinghe}, O. and {Amato}, A. and {Amra}, C. and {Ananyeva}, A. and {Anderson}, S.~B. and {Anderson}, W.~G. and {Andia}, M. and {Ando}, M. and {Andrade}, T. and {Andr{\'e}s-Carcasona}, M. and {Andri{\'c}}, T. and {Anglin}, J. and {Ansoldi}, S. and {Antelis}, J.~M. and {Antier}, S. and {Aoumi}, M. and {Appavuravther}, E.~Z. and {Appert}, S. and {Apple}, S.~K. and {Arai}, K. and {Araya}, A. and {Araya}, M.~C. and {Arca Sedda}, M. and {Areeda}, J.~S. and {Argianas}, L. and {Aritomi}, N. and {Armato}, F. and {Armstrong}, S. and {Arnaud}, N. and {Arogeti}, M. and {Aronson}, S.~M. and {Arun}, K.~G. and {Ashton}, G. and {Aso}, Y. and {Assiduo}, M. and {Assis de Souza Melo}, S. and {Aston}, S.~M. and {Astone}, P. and {Attadio}, F. and {Aubin}, F. and {Aultoneal}, K. and {Avallone}, G. and {Babak}, S. and {Badaracco}, F. and {Badger}, C. and {Bae}, S. and {Bagnasco}, S. and {Bagui}, E. and {Baiotti}, L. and {Bajpai}, R. and {Baka}, T. and {Baker}, T. and {Ball}, M. and {Ballardin}, G. and {Ballmer}, S.~W. and {Banagiri}, S. and {Banerjee}, B. and {Bankar}, D. and {Baptiste}, T.~M. and {Baral}, P. and {Barayoga}, J.~C. and {Barish}, B.~C. and {Barker}, D. and {Barman}, N. and {Barneo}, P. and {Barone}, F. and {Barr}, B. and {Barsotti}, L. and {Barsuglia}, M. and {Barta}, D. and {Bartoletti}, A.~M. and {Barton}, M.~A. and {Bartos}, I. and {Basak}, S. and {Basalaev}, A. and {Bassiri}, R. and {Basti}, A. and {Bates}, D.~E. and {Bawaj}, M. and {Baxi}, P. and {Bayley}, J.~C. and {Baylor}, A.~C. and {Baynard}, II, P.~A. and {Bazzan}, M. and {Bedakihale}, V.~M. and {Beirnaert}, F. and {Bejger}, M. and {Belardinelli}, D. and {Bell}, A.~S. and {Bellie}, D.~S. and {Bellizzi}, L. and {Beltran-Martinez}, D. and {Benoit}, W. and {Bentara}, I. and {Bentley}, J.~D. and {Ben Yaala}, M. and {Bera}, S. and {Bergamin}, F. and {Berger}, B.~K. and {Bernuzzi}, S. and {Beroiz}, M. and {Berry}, C.~P.~L. and {Bersanetti}, D. and {Bertolini}, A. and {Betzwieser}, J. and {Beveridge}, D. and {Bevilacqua}, G. and {Bevins}, N. and {Bhandare}, R. and {Bhatt}, R. and {Bhattacharjee}, D. and {Bhaumik}, S. and {Bhowmick}, S. and {Biancalana}, V. and {Bianchi}, A. and {Bilenko}, I.~A. and {Billingsley}, G. and {Binetti}, A. and {Bini}, S. and {Binu}, C. and {Birnholtz}, O. and {Biscoveanu}, S. and {Bisht}, A. and {Bitossi}, M. and {Bizouard}, M.-A. and {Blaber}, S. and {Blackburn}, J.~K. and {Blagg}, L.~A. and {Blair}, C.~D. and {Blair}, D.~G. and {Bobba}, F. and {Bode}, N. and {Boileau}, G. and {Boldrini}, M. and {Bolingbroke}, G.~N. and {Bolliand}, A. and {Bonavena}, L.~D. and {Bondarescu}, R. and {Bondu}, F. and {Bonilla}, E. and {Bonilla}, M.~S. and {Bonino}, A. and {Bonnand}, R. and {Booker}, P. and {Borchers}, A. and {Borhanian}, S. and {Boschi}, V. and {Bose}, S. and {Bossilkov}, V. and {Boudon}, A. and {Bozzi}, A. and {Bradaschia}, C. and {Brady}, P.~R. and {Branch}, A. and {Branchesi}, M. and {Braun}, I. and {Briant}, T. and {Brillet}, A. and {Brinkmann}, M. and {Brockill}, P. and {Brockmueller}, E. and {Brooks}, A.~F. and {Brown}, B.~C. and {Brown}, D.~D. and {Brozzetti}, M.~L. and {Brunett}, S. and {Bruno}, G. and {Bruntz}, R. and {Bryant}, J. and {Bu}, Y. and {Bucci}, F. and {Buchanan}, J.},
        title = "{GWTC-4.0: Population Properties of Merging Compact Binaries}",
      journal = {\apjl},
         year = 2026,
        month = jul,
       volume = {1005},
       number = {2},
          eid = {L51},
        pages = {L51},
          doi = {10.3847/2041-8213/ae771e},
archivePrefix = {arXiv},
       eprint = {2508.18083},
 primaryClass = {astro-ph.HE},
       adsurl = {https://ui.adsabs.harvard.edu/abs/2026ApJ..1005L..51A}
}

@ARTICLE{McK14,
       author = {{McKernan}, B. and {Ford}, K.~E.~S. and {Kocsis}, B. and {Lyra}, W. and {Winter}, L.~M.},
        title = "{Intermediate-mass black holes in AGN discs - II. Model predictions and observational constraints}",
      journal = {\mnras},
         year = 2014,
        month = jun,
       volume = {441},
       number = {1},
        pages = {900-909},
          doi = {10.1093/mnras/stu553},
archivePrefix = {arXiv},
       eprint = {1403.6433},
 primaryClass = {astro-ph.GA},
       adsurl = {https://ui.adsabs.harvard.edu/abs/2014MNRAS.441..900M}
}

@article{Thompson2005,
	author = {{Thompson}, T.~A. and {Quataert}, E. and {Murray}, N.},
	journal = {\apj},
	month = sep,
	pages = {167-185},
	title = {{Radiation Pressure-supported Starburst Disks and Active Galactic Nucleus Fueling}},
	volume = 630,
	year = 2005}

@ARTICLE{Tanaka2004,
       author = {{Tanaka}, Hidekazu and {Ward}, William R.},
        title = "{Three-dimensional Interaction between a Planet and an Isothermal Gaseous Disk. II. Eccentricity Waves and Bending Waves}",
      journal = {\apj},
         year = 2004,
        month = feb,
       volume = {602},
       number = {1},
        pages = {388-395},
          doi = {10.1086/380992},
       adsurl = {https://ui.adsabs.harvard.edu/abs/2004ApJ...602..388T}
}

@ARTICLE{Ward1997,
       author = {{Ward}, William R.},
        title = "{Survival of Planetary Systems}",
      journal = {\apjl},
         year = 1997,
        month = jun,
       volume = {482},
       number = {2},
        pages = {L211-L214},
          doi = {10.1086/310701},
       adsurl = {https://ui.adsabs.harvard.edu/abs/1997ApJ...482L.211W}
}

@ARTICLE{Goldreich1979,
       author = {{Goldreich}, P. and {Tremaine}, S.},
        title = "{The excitation of density waves at the Lindblad and corotation resonances by an external potential.}",
      journal = {\apj},
         year = 1979,
        month = nov,
       volume = {233},
        pages = {857-871},
          doi = {10.1086/157448},
       adsurl = {https://ui.adsabs.harvard.edu/abs/1979ApJ...233..857G}
}

@ARTICLE{Haehnelt:1993,
       author = {{Haehnelt}, Martin G. and {Rees}, Martin J.},
        title = "{The formation of nuclei in newly formed galaxies and the evolution of the quasar population}",
      journal = {\mnras},
         year = 1993,
        month = jul,
       volume = {263},
       number = {1},
        pages = {168-178},
          doi = {10.1093/mnras/263.1.168},
       adsurl = {https://ui.adsabs.harvard.edu/abs/1993MNRAS.263..168H}
}

@ARTICLE{Abbott:2023,
       author = {{Abbott}, R. and {Abbott}, T.~D. and {Acernese}, F. and {Ackley}, K. and {Adams}, C. and {Adhikari}, N. and {Adhikari}, R.~X. and {Adya}, V.~B. and {Affeldt}, C. and {Agarwal}, D. and {Agathos}, M. and {Agatsuma}, K. and {Aggarwal}, N. and {Aguiar}, O.~D. and {Aiello}, L. and {Ain}, A. and {Ajith}, P. and {Akcay}, S. and {Akutsu}, T. and {Albanesi}, S. and {Allocca}, A. and {Altin}, P.~A. and {Amato}, A. and {Anand}, C. and {Anand}, S. and {Ananyeva}, A. and {Anderson}, S.~B. and {Anderson}, W.~G. and {Ando}, M. and {Andrade}, T. and {Andres}, N. and {Andri{\'c}}, T. and {Angelova}, S.~V. and {Ansoldi}, S. and {Antelis}, J.~M. and {Antier}, S. and {Appert}, S. and {Arai}, Koji and {Arai}, Koya and {Arai}, Y. and {Araki}, S. and {Araya}, A. and {Araya}, M.~C. and {Areeda}, J.~S. and {Ar{\`e}ne}, M. and {Aritomi}, N. and {Arnaud}, N. and {Arogeti}, M. and {Aronson}, S.~M. and {Arun}, K.~G. and {Asada}, H. and {Asali}, Y. and {Ashton}, G. and {Aso}, Y. and {Assiduo}, M. and {Aston}, S.~M. and {Astone}, P. and {Aubin}, F. and {Austin}, C. and {Babak}, S. and {Badaracco}, F. and {Bader}, M.~K.~M. and {Badger}, C. and {Bae}, S. and {Bae}, Y. and {Baer}, A.~M. and {Bagnasco}, S. and {Bai}, Y. and {Baiotti}, L. and {Baird}, J. and {Bajpai}, R. and {Ball}, M. and {Ballardin}, G. and {Ballmer}, S.~W. and {Balsamo}, A. and {Baltus}, G. and {Banagiri}, S. and {Bankar}, D. and {Barayoga}, J.~C. and {Barbieri}, C. and {Barish}, B.~C. and {Barker}, D. and {Barneo}, P. and {Barone}, F. and {Barr}, B. and {Barsotti}, L. and {Barsuglia}, M. and {Barta}, D. and {Bartlett}, J. and {Barton}, M.~A. and {Bartos}, I. and {Bassiri}, R. and {Basti}, A. and {Bawaj}, M. and {Bayley}, J.~C. and {Baylor}, A.~C. and {Bazzan}, M. and {B{\'e}csy}, B. and {Bedakihale}, V.~M. and {Bejger}, M. and {Belahcene}, I. and {Benedetto}, V. and {Beniwal}, D. and {Bennett}, T.~F. and {Bentley}, J.~D. and {Benyaala}, M. and {Bergamin}, F. and {Berger}, B.~K. and {Bernuzzi}, S. and {Berry}, C.~P.~L. and {Bersanetti}, D. and {Bertolini}, A. and {Betzwieser}, J. and {Beveridge}, D. and {Bhandare}, R. and {Bhardwaj}, U. and {Bhattacharjee}, D. and {Bhaumik}, S. and {Bilenko}, I.~A. and {Billingsley}, G. and {Bini}, S. and {Birney}, R. and {Birnholtz}, O. and {Biscans}, S. and {Bischi}, M. and {Biscoveanu}, S. and {Bisht}, A. and {Biswas}, B. and {Bitossi}, M. and {Bizouard}, M.-A. and {Blackburn}, J.~K. and {Blair}, C.~D. and {Blair}, D.~G. and {Blair}, R.~M. and {Bobba}, F. and {Bode}, N. and {Boer}, M. and {Bogaert}, G. and {Boldrini}, M. and {Bonavena}, L.~D. and {Bondu}, F. and {Bonilla}, E. and {Bonnand}, R. and {Booker}, P. and {Boom}, B.~A. and {Bork}, R. and {Boschi}, V. and {Bose}, N. and {Bose}, S. and {Bossilkov}, V. and {Boudart}, V. and {Bouffanais}, Y. and {Bozzi}, A. and {Bradaschia}, C. and {Brady}, P.~R. and {Bramley}, A. and {Branch}, A. and {Branchesi}, M. and {Brandt}, J. and {Brau}, J.~E. and {Breschi}, M. and {Briant}, T. and {Briggs}, J.~H. and {Brillet}, A. and {Brinkmann}, M. and {Brockill}, P. and {Brooks}, A.~F. and {Brooks}, J. and {Brown}, D.~D. and {Brunett}, S. and {Bruno}, G. and {Bruntz}, R. and {Bryant}, J. and {Bulik}, T. and {Bulten}, H.~J. and {Buonanno}, A. and {Buscicchio}, R. and {Buskulic}, D. and {Buy}, C. and {Byer}, R.~L. and {Davies}, G.~S. Cabourn and {Cadonati}, L. and {Cagnoli}, G. and {Cahillane}, C. and {Bustillo}, J. Calder{\'o}n and {Callaghan}, J.~D. and {Callister}, T.~A. and {Calloni}, E. and {Cameron}, J. and {Camp}, J.~B. and {Canepa}, M. and {Canevarolo}, S. and {Cannavacciuolo}, M. and {Cannon}, K.~C. and {Cao}, H. and {Cao}, Z. and {Capocasa}, E. and {Capote}, E. and {Carapella}, G. and {Carbognani}, F.},
        title = "{GWTC-3: Compact Binary Coalescences Observed by LIGO and Virgo during the Second Part of the Third Observing Run}",
      journal = {Physical Review X},
         year = 2023,
        month = oct,
       volume = {13},
       number = {4},
          eid = {041039},
        pages = {041039},
          doi = {10.1103/PhysRevX.13.041039},
archivePrefix = {arXiv},
       eprint = {2111.03606},
 primaryClass = {gr-qc},
       adsurl = {https://ui.adsabs.harvard.edu/abs/2023PhRvX..13d1039A}
}

@ARTICLE{Marconi:2004,
       author = {{Marconi}, A. and {Risaliti}, G. and {Gilli}, R. and {Hunt}, L.~K. and {Maiolino}, R. and {Salvati}, M.},
        title = "{Local supermassive black holes, relics of active galactic nuclei and the X-ray background}",
      journal = {\mnras},
         year = 2004,
        month = jun,
       volume = {351},
       number = {1},
        pages = {169-185},
          doi = {10.1111/j.1365-2966.2004.07765.x},
archivePrefix = {arXiv},
       eprint = {astro-ph/0311619},
 primaryClass = {astro-ph},
       adsurl = {https://ui.adsabs.harvard.edu/abs/2004MNRAS.351..169M}
}

@ARTICLE{Dodici24,
       author = {{Dodici}, Mark and {Tremaine}, Scott},
        title = "{Studying Binary Formation under Dynamical Friction Using Hill's Problem}",
      journal = {\apj},
         year = 2024,
        month = sep,
       volume = {972},
       number = {2},
          eid = {193},
        pages = {193},
          doi = {10.3847/1538-4357/ad5cf2},
archivePrefix = {arXiv},
       eprint = {2404.08138},
 primaryClass = {astro-ph.GA},
       adsurl = {https://ui.adsabs.harvard.edu/abs/2024ApJ...972..193D}
}

@ARTICLE{LISA23,
       author = {{Amaro-Seoane}, Pau and {Andrews}, Jeff and {Arca Sedda}, Manuel and {Askar}, Abbas and {Baghi}, Quentin and {Balasov}, Razvan and {Bartos}, Imre and {Bavera}, Simone S. and {Bellovary}, Jillian and {Berry}, Christopher P.~L. and {Berti}, Emanuele and {Bianchi}, Stefano and {Blecha}, Laura and {Blondin}, St{\'e}phane and {Bogdanovi{\'c}}, Tamara and {Boissier}, Samuel and {Bonetti}, Matteo and {Bonoli}, Silvia and {Bortolas}, Elisa and {Breivik}, Katelyn and {Capelo}, Pedro R. and {Caramete}, Laurentiu and {Cattorini}, Federico and {Charisi}, Maria and {Chaty}, Sylvain and {Chen}, Xian and {Chru{\'s}li{\'n}ska}, Martyna and {Chua}, Alvin J.~K. and {Church}, Ross and {Colpi}, Monica and {D'Orazio}, Daniel and {Danielski}, Camilla and {Davies}, Melvyn B. and {Dayal}, Pratika and {De Rosa}, Alessandra and {Derdzinski}, Andrea and {Destounis}, Kyriakos and {Dotti}, Massimo and {Du{\c{t}}an}, Ioana and {Dvorkin}, Irina and {Fabj}, Gaia and {Foglizzo}, Thierry and {Ford}, Saavik and {Fouvry}, Jean-Baptiste and {Franchini}, Alessia and {Fragos}, Tassos and {Fryer}, Chris and {Gaspari}, Massimo and {Gerosa}, Davide and {Graziani}, Luca and {Groot}, Paul and {Habouzit}, Melanie and {Haggard}, Daryl and {Haiman}, Zoltan and {Han}, Wen-Biao and {Istrate}, Alina and {Johansson}, Peter H. and {Khan}, Fazeel Mahmood and {Kimpson}, Tomas and {Kokkotas}, Kostas and {Kong}, Albert and {Korol}, Valeriya and {Kremer}, Kyle and {Kupfer}, Thomas and {Lamberts}, Astrid and {Larson}, Shane and {Lau}, Mike and {Liu}, Dongliang and {Lloyd-Ronning}, Nicole and {Lodato}, Giuseppe and {Lupi}, Alessandro and {Ma}, Chung-Pei and {Maccarone}, Tomas and {Mandel}, Ilya and {Mangiagli}, Alberto and {Mapelli}, Michela and {Mathis}, St{\'e}phane and {Mayer}, Lucio and {McGee}, Sean and {McKernan}, Berry and {Miller}, M. Coleman and {Mota}, David F. and {Mumpower}, Matthew and {Nasim}, Syeda S. and {Nelemans}, Gijs and {Noble}, Scott and {Pacucci}, Fabio and {Panessa}, Francesca and {Paschalidis}, Vasileios and {Pfister}, Hugo and {Porquet}, Delphine and {Quenby}, John and {Ricarte}, Angelo and {R{\"o}pke}, Friedrich K. and {Regan}, John and {Rosswog}, Stephan and {Ruiter}, Ashley and {Ruiz}, Milton and {Runnoe}, Jessie and {Schneider}, Raffaella and {Schnittman}, Jeremy and {Secunda}, Amy and {Sesana}, Alberto and {Seto}, Naoki and {Shao}, Lijing and {Shapiro}, Stuart and {Sopuerta}, Carlos and {Stone}, Nicholas C. and {Suvorov}, Arthur and {Tamanini}, Nicola and {Tamfal}, Tomas and {Tauris}, Thomas and {Temmink}, Karel and {Tomsick}, John and {Toonen}, Silvia and {Torres-Orjuela}, Alejandro and {Toscani}, Martina and {Tsokaros}, Antonios and {Unal}, Caner and {V{\'a}zquez-Aceves}, Ver{\'o}nica and {Valiante}, Rosa and {van Putten}, Maurice and {van Roestel}, Jan and {Vignali}, Christian and {Volonteri}, Marta and {Wu}, Kinwah and {Younsi}, Ziri and {Yu}, Shenghua and {Zane}, Silvia and {Zwick}, Lorenz and {Antonini}, Fabio and {Baibhav}, Vishal and {Barausse}, Enrico and {Bonilla Rivera}, Alexander and {Branchesi}, Marica and {Branduardi-Raymont}, Graziella and {Burdge}, Kevin and {Chakraborty}, Srija and {Cuadra}, Jorge and {Dage}, Kristen and {Davis}, Benjamin and {de Mink}, Selma E. and {Decarli}, Roberto and {Doneva}, Daniela and {Escoffier}, Stephanie and {Gandhi}, Poshak and {Haardt}, Francesco and {Lousto}, Carlos O. and {Nissanke}, Samaya and {Nordhaus}, Jason and {O'Shaughnessy}, Richard and {Portegies Zwart}, Simon and {Pound}, Adam and {Schussler}, Fabian and {Sergijenko}, Olga and {Spallicci}, Alessandro and {Vernieri}, Daniele and {Vigna-G{\'o}mez}, Alejandro},
        title = "{Astrophysics with the Laser Interferometer Space Antenna}",
      journal = {Living Reviews in Relativity},
         year = 2023,
        month = dec,
       volume = {26},
       number = {1},
          eid = {2},
        pages = {2},
          doi = {10.1007/s41114-022-00041-y},
archivePrefix = {arXiv},
       eprint = {2203.06016},
 primaryClass = {gr-qc},
       adsurl = {https://ui.adsabs.harvard.edu/abs/2023LRR....26....2A}
}

@ARTICLE{RomSari25,
       author = {{Rom}, Barak and {Sari}, Re'em},
        title = "{Mass Segregation and Transient Formation in Nuclear Stellar Clusters}",
      journal = {\apj},
         year = 2025,
        month = oct,
       volume = {991},
       number = {2},
          eid = {146},
        pages = {146},
          doi = {10.3847/1538-4357/adfb6c},
archivePrefix = {arXiv},
       eprint = {2502.13209},
 primaryClass = {astro-ph.HE},
       adsurl = {https://ui.adsabs.harvard.edu/abs/2025ApJ...991..146R}
}

@ARTICLE{Secunda:2021,
       author = {{Secunda}, Amy and {Hernandez}, Betsy and {Goodman}, Jeremy and {Leigh}, Nathan W.~C. and {McKernan}, Barry and {Ford}, K.~E. Saavik and {Adorno}, Jose I.},
        title = "{Evolution of Retrograde Orbiters in an Active Galactic Nucleus Disk}",
      journal = {\apjl},
         year = 2021,
        month = feb,
       volume = {908},
       number = {2},
          eid = {L27},
        pages = {L27},
          doi = {10.3847/2041-8213/abe11d},
archivePrefix = {arXiv},
       eprint = {2009.03910},
 primaryClass = {astro-ph.HE},
       adsurl = {https://ui.adsabs.harvard.edu/abs/2021ApJ...908L..27S}
}

@ARTICLE{Delfavero25b,
       author = {{Delfavero}, V. and {Ray}, S. and {Cook}, H.~E. and {Nathaniel}, K. and {McKernan}, B. and {Ford}, K.~E.~S. and {Postiglione}, J. and {McPike}, E. and {O'Shaughnessy}, R.},
        title = "{Prospects for the formation of GW231123 from the AGN channel}",
      journal = {arXiv e-prints},
         year = 2025,
        month = aug,
          eid = {arXiv:2508.13412},
        pages = {arXiv:2508.13412},
          doi = {10.48550/arXiv.2508.13412},
archivePrefix = {arXiv},
       eprint = {2508.13412},
 primaryClass = {gr-qc},
       adsurl = {https://ui.adsabs.harvard.edu/abs/2025arXiv250813412D}
}

@ARTICLE{Roy25,
       author = {{Kishore Roy}, Soumendra and {van Son}, Lieke A.~C. and {Farr}, Will M.},
        title = "{A mid-thirties crisis: dissecting the properties of gravitational wave sources near the 35 solar mass peak}",
      journal = {Classical and Quantum Gravity},
         year = 2025,
        month = nov,
       volume = {42},
       number = {22},
          eid = {225008},
        pages = {225008},
          doi = {10.1088/1361-6382/ae1921},
archivePrefix = {arXiv},
       eprint = {2507.01086},
 primaryClass = {astro-ph.HE},
       adsurl = {https://ui.adsabs.harvard.edu/abs/2025CQGra..42v5008K}
}

@ARTICLE{RixinLi22,
       author = {{Li}, Rixin and {Lai}, Dong},
        title = "{Hydrodynamical evolution of black-hole binaries embedded in AGN discs}",
      journal = {\mnras},
         year = 2022,
        month = dec,
       volume = {517},
       number = {2},
        pages = {1602-1624},
          doi = {10.1093/mnras/stac2577},
archivePrefix = {arXiv},
       eprint = {2202.07633},
 primaryClass = {astro-ph.HE},
       adsurl = {https://ui.adsabs.harvard.edu/abs/2022MNRAS.517.1602L}
}

@ARTICLE{YaPingLi21,
       author = {{Li}, Ya-Ping and {Dempsey}, Adam M. and {Li}, Shengtai and {Li}, Hui and {Li}, Jiaru},
        title = "{Orbital Evolution of Binary Black Holes in Active Galactic Nucleus Disks: A Disk Channel for Binary Black Hole Mergers?}",
      journal = {\apj},
         year = 2021,
        month = apr,
       volume = {911},
       number = {2},
          eid = {124},
        pages = {124},
          doi = {10.3847/1538-4357/abed48},
archivePrefix = {arXiv},
       eprint = {2101.09406},
 primaryClass = {astro-ph.HE},
       adsurl = {https://ui.adsabs.harvard.edu/abs/2021ApJ...911..124L}
}

@ARTICLE{Jairu23,
       author = {{Li}, Jiaru and {Dempsey}, Adam M. and {Li}, Hui and {Lai}, Dong and {Li}, Shengtai},
        title = "{Hydrodynamical Simulations of Black Hole Binary Formation in AGN Disks}",
      journal = {\apjl},
         year = 2023,
        month = feb,
       volume = {944},
       number = {2},
          eid = {L42},
        pages = {L42},
          doi = {10.3847/2041-8213/acb934},
archivePrefix = {arXiv},
       eprint = {2211.10357},
 primaryClass = {astro-ph.HE},
       adsurl = {https://ui.adsabs.harvard.edu/abs/2023ApJ...944L..42L}
}

@ARTICLE{Whitehead25,
       author = {{Whitehead}, Henry and {Rowan}, Connar and {Kocsis}, Bence},
        title = "{3D adiabatic simulations of binary black hole formation in AGN discs}",
      journal = {\mnras},
         year = 2025,
        month = sep,
       volume = {542},
       number = {2},
        pages = {1033-1055},
          doi = {10.1093/mnras/staf1271},
archivePrefix = {arXiv},
       eprint = {2502.14959},
 primaryClass = {astro-ph.HE},
       adsurl = {https://ui.adsabs.harvard.edu/abs/2025MNRAS.542.1033W}
}

@ARTICLE{Tagawa20,
       author = {{Tagawa}, Hiromichi and {Haiman}, Zolt{\'a}n and {Kocsis}, Bence},
        title = "{Formation and Evolution of Compact-object Binaries in AGN Disks}",
      journal = {\apj},
         year = 2020,
        month = jul,
       volume = {898},
       number = {1},
          eid = {25},
        pages = {25},
          doi = {10.3847/1538-4357/ab9b8c},
archivePrefix = {arXiv},
       eprint = {1912.08218},
 primaryClass = {astro-ph.GA},
       adsurl = {https://ui.adsabs.harvard.edu/abs/2020ApJ...898...25T}
}

@ARTICLE{Delfavero25,
       author = {{Delfavero}, Vera and {Ford}, K.~E. Saavik and {McKernan}, Barry and {Cook}, Harrison E. and {Nathaniel}, Kaila and {Postiglione}, Jake and {Ray}, Shawn and {McPike}, Emily and {O'Shaughnessy}, Richard},
        title = "{McFACTS III: Compact Binary Mergers from Active Galactic Nucleus Disks over an Entire Synthetic Universe}",
      journal = {\apj},
         year = 2025,
        month = aug,
       volume = {989},
       number = {1},
          eid = {67},
        pages = {67},
          doi = {10.3847/1538-4357/ade4c1},
archivePrefix = {arXiv},
       eprint = {2410.18815},
 primaryClass = {astro-ph.HE},
       adsurl = {https://ui.adsabs.harvard.edu/abs/2025ApJ...989...67D}
}

@ARTICLE{Pan21,
       author = {{Pan}, Zhen and {Yang}, Huan},
        title = "{Formation rate of extreme mass ratio inspirals in active galactic nuclei}",
      journal = {\prd},
         year = 2021,
        month = may,
       volume = {103},
       number = {10},
          eid = {103018},
        pages = {103018},
          doi = {10.1103/PhysRevD.103.103018},
archivePrefix = {arXiv},
       eprint = {2101.09146},
 primaryClass = {astro-ph.HE},
       adsurl = {https://ui.adsabs.harvard.edu/abs/2021PhRvD.103j3018P}
}

@ARTICLE{PengXian21,
       author = {{Peng}, Peng and {Chen}, Xian},
        title = "{The last migration trap of compact objects in AGN accretion disc}",
      journal = {\mnras},
         year = 2021,
        month = jul,
       volume = {505},
       number = {1},
        pages = {1324-1333},
          doi = {10.1093/mnras/stab1419},
archivePrefix = {arXiv},
       eprint = {2104.07685},
 primaryClass = {astro-ph.HE},
       adsurl = {https://ui.adsabs.harvard.edu/abs/2021MNRAS.505.1324P}
}

@ARTICLE{Yang19,
       author = {{Yang}, Y. and {Bartos}, I. and {Gayathri}, V. and {Ford}, K.~E.~S. and {Haiman}, Z. and {Klimenko}, S. and {Kocsis}, B. and {M{\'a}rka}, S. and {M{\'a}rka}, Z. and {McKernan}, B. and {O'Shaughnessy}, R.},
        title = "{Hierarchical Black Hole Mergers in Active Galactic Nuclei}",
      journal = {\prl},
         year = 2019,
        month = nov,
       volume = {123},
       number = {18},
          eid = {181101},
        pages = {181101},
          doi = {10.1103/PhysRevLett.123.181101},
archivePrefix = {arXiv},
       eprint = {1906.09281},
 primaryClass = {astro-ph.HE},
       adsurl = {https://ui.adsabs.harvard.edu/abs/2019PhRvL.123r1101Y}
}

@ARTICLE{McK12,
       author = {{McKernan}, B. and {Ford}, K.~E.~S. and {Lyra}, W. and {Perets}, H.~B.},
        title = "{Intermediate mass black holes in AGN discs - I. Production and growth}",
      journal = {\mnras},
         year = 2012,
        month = sep,
       volume = {425},
       number = {1},
        pages = {460-469},
          doi = {10.1111/j.1365-2966.2012.21486.x},
archivePrefix = {arXiv},
       eprint = {1206.2309},
 primaryClass = {astro-ph.GA},
       adsurl = {https://ui.adsabs.harvard.edu/abs/2012MNRAS.425..460M}
}

@ARTICLE{Grishin24,
       author = {{Grishin}, Evgeni and {Gilbaum}, Shmuel and {Stone}, Nicholas C.},
        title = "{The effect of thermal torques on AGN disc migration traps and gravitational wave populations}",
      journal = {\mnras},
         year = 2024,
        month = may,
       volume = {530},
       number = {2},
        pages = {2114-2132},
          doi = {10.1093/mnras/stae828},
archivePrefix = {arXiv},
       eprint = {2307.07546},
 primaryClass = {astro-ph.HE},
       adsurl = {https://ui.adsabs.harvard.edu/abs/2024MNRAS.530.2114G}
}

@ARTICLE{Vaccaro25,
       author = {{Vaccaro}, Maria Paola and {Seif}, Yannick and {Mapelli}, Michela},
        title = "{The role of migration traps in the formation of binary black holes in AGN disks}",
      journal = {arXiv e-prints},
         year = 2025,
        month = aug,
          eid = {arXiv:2508.03637},
        pages = {arXiv:2508.03637},
          doi = {10.48550/arXiv.2508.03637},
archivePrefix = {arXiv},
       eprint = {2508.03637},
 primaryClass = {astro-ph.HE},
       adsurl = {https://ui.adsabs.harvard.edu/abs/2025arXiv250803637V}
}

@ARTICLE{FM25,
       author = {{Ford}, K.~E. Saavik and {McKernan}, Barry},
        title = "{Using gravitational waves and multi-messenger Astronomy to reverse-engineer the properties of galactic nuclei}",
      journal = {arXiv e-prints},
         year = 2025,
        month = jun,
          eid = {arXiv:2506.08801},
        pages = {arXiv:2506.08801},
          doi = {10.48550/arXiv.2506.08801},
archivePrefix = {arXiv},
       eprint = {2506.08801},
 primaryClass = {astro-ph.HE},
       adsurl = {https://ui.adsabs.harvard.edu/abs/2025arXiv250608801F}
}

@ARTICLE{King:2015,
       author = {{King}, Andrew and {Nixon}, Chris},
        title = "{AGN flickering and chaotic accretion}",
      journal = {\mnras},
         year = 2015,
        month = oct,
       volume = {453},
       number = {1},
        pages = {L46-L47},
          doi = {10.1093/mnrasl/slv098},
archivePrefix = {arXiv},
       eprint = {1507.05960},
 primaryClass = {astro-ph.HE},
       adsurl = {https://ui.adsabs.harvard.edu/abs/2015MNRAS.453L..46K}
}

@ARTICLE{Tagawa26,
       author = {{Tagawa}, Hiromichi and {Haiman}, Zolt{\'a}n and {Kocsis}, Bence},
        title = "{Properties of Black Hole Mergers in Disks of Active Galactic Nuclei}",
      journal = {\apj},
         year = 2026,
        month = aug,
       volume = {1007},
       number = {1},
          eid = {67},
        pages = {67},
          doi = {10.3847/1538-4357/ae8760},
       adsurl = {https://ui.adsabs.harvard.edu/abs/2026ApJ..1007...67T}
}

@ARTICLE{Moncrieff26,
       author = {{Moncrieff}, Jordan W.~N. and {Grishin}, Evgeni and {Trani}, Alessandro A. and {Panther}, Fiona H. and {Pietrosanti}, Olga},
        title = "{Not all roads lead to merger: AGN disc properties influence the interactions of highly unequal mass black holes}",
      journal = {\mnras},
         year = 2026,
        month = jan,
       volume = {545},
       number = {3},
          eid = {staf2217},
        pages = {staf2217},
          doi = {10.1093/mnras/staf2217},
archivePrefix = {arXiv},
       eprint = {2511.09129},
 primaryClass = {astro-ph.HE},
       adsurl = {https://ui.adsabs.harvard.edu/abs/2026MNRAS.545f2217M}
}

@ARTICLE{GWTC4a,
       author = {{Abac}, A.~G. and {Abouelfettouh}, I. and {Acernese}, F. and {Ackley}, K. and {Adamcewicz}, C. and {Adhicary}, S. and {Adhikari}, D. and {Adhikari}, N. and {Adhikari}, R.~X. and {Adkins}, V.~K. and {Afroz}, S. and {Agapito}, A. and {Agarwal}, D. and {Agathos}, M. and {Aggarwal}, N. and {Aggarwal}, S. and {Aguiar}, O.~D. and {Ahrend}, I.-L. and {Aiello}, L. and {Ain}, A. and {Ajith}, P. and {Akcay}, S. and {Akutsu}, T. and {Albanesi}, S. and {Ali}, W. and {Al-Kershi}, S. and {All{\'e}n{\'e}}, C. and {Allocca}, A. and {Al-Shammari}, S. and {Altin}, P.~A. and {Alvarez-Lopez}, S. and {Amar}, W. and {Amarasinghe}, O. and {Amato}, A. and {Amicucci}, F. and {Amra}, C. and {Ananyeva}, A. and {Anderson}, S.~B. and {Anderson}, W.~G. and {Andia}, M. and {Ando}, M. and {Andr{\'e}s-Carcasona}, M. and {Andri{\'c}}, T. and {Anglin}, J. and {Ansoldi}, S. and {Antelis}, J.~M. and {Antier}, S. and {Aoumi}, M. and {Appavuravther}, E.~Z. and {Appert}, S. and {Apple}, S.~K. and {Arai}, K. and {Araya}, A. and {Araya}, M.~C. and {Arca Sedda}, M. and {Areeda}, J.~S. and {Aritomi}, N. and {Armato}, F. and {Armstrong}, S. and {Arnaud}, N. and {Arogeti}, M. and {Aronson}, S.~M. and {Arun}, K.~G. and {Ashton}, G. and {Aso}, Y. and {Asprea}, L. and {Assiduo}, M. and {Assis de Souza Melo}, S. and {Aston}, S.~M. and {Astone}, P. and {Attadio}, F. and {Aubin}, F. and {Aultoneal}, K. and {Avallone}, G. and {Avila}, E.~A. and {Babak}, S. and {Badger}, C. and {Bae}, S. and {Bagnasco}, S. and {Baiotti}, L. and {Bajpai}, R. and {Baka}, T. and {Baker}, A.~M. and {Baker}, K.~A. and {Baker}, T. and {Baldi}, G. and {Baldicchi}, N. and {Ball}, M. and {Ballardin}, G. and {Ballmer}, S.~W. and {Banagiri}, S. and {Banerjee}, B. and {Bankar}, D. and {Baptiste}, T.~M. and {Baral}, P. and {Baratti}, M. and {Barayoga}, J.~C. and {Barish}, B.~C. and {Barker}, D. and {Barman}, N. and {Barneo}, P. and {Barone}, F. and {Barr}, B. and {Barsotti}, L. and {Barsuglia}, M. and {Barta}, D. and {Bartoletti}, A.~M. and {Barton}, M.~A. and {Bartos}, I. and {Basalaev}, A. and {Bassiri}, R. and {Basti}, A. and {Bawaj}, M. and {Baxi}, P. and {Bayley}, J.~C. and {Baylor}, A.~C. and {Baynard}, II, P.~A. and {Bazzan}, M. and {Bedakihale}, V.~M. and {Beirnaert}, F. and {Bejger}, M. and {Belardinelli}, D. and {Bell}, A.~S. and {Bellie}, D.~S. and {Bellizzi}, L. and {Benoit}, W. and {Bentara}, I. and {Bentley}, J.~D. and {Ben Yaala}, M. and {Bera}, S. and {Bergamin}, F. and {Berger}, B.~K. and {Bernuzzi}, S. and {Beroiz}, M. and {Berry}, C.~P.~L. and {Bersanetti}, D. and {Bertheas}, T. and {Bertolini}, A. and {Betzwieser}, J. and {Beveridge}, D. and {Bevilacqua}, G. and {Bevins}, N. and {Bhandare}, R. and {Bhat}, S.~A. and {Bhatt}, R. and {Bhattacharjee}, D. and {Bhattacharyya}, S. and {Bhaumik}, S. and {Biancalana}, V. and {Bianchi}, A. and {Bilenko}, I.~A. and {Billingsley}, G. and {Binetti}, A. and {Bini}, S. and {Binu}, C. and {Biot}, S. and {Birnholtz}, O. and {Biscoveanu}, S. and {Bisht}, A. and {Bitossi}, M. and {Bizouard}, M.-A. and {Blaber}, S. and {Blackburn}, J.~K. and {Blagg}, L.~A. and {Blair}, C.~D. and {Blair}, D.~G. and {Bode}, N. and {Boettner}, N. and {Boileau}, G. and {Boldrini}, M. and {Bolingbroke}, G.~N. and {Bolliand}, A. and {Bonavena}, L.~D. and {Bondarescu}, R. and {Bondu}, F. and {Bonilla}, E. and {Bonilla}, M.~S. and {Bonino}, A. and {Bonnand}, R. and {Borchers}, A. and {Boschi}, V. and {Bose}, S. and {Bossilkov}, V. and {Bothra}, Y. and {Boudon}, A. and {Bourg}, L. and {Boyle}, M. and {Bozzi}, A. and {Bradaschia}, C. and {Brady}, P.~R. and {Branch}, A. and {Branchesi}, M. and {Braun}, I. and {Briant}, T. and {Brillet}, A. and {Brinkmann}, M. and {Brockill}, P. and {Brockmueller}, E. and {Brooks}, A.~F. and {Brown}, B.~C.},
        title = "{GWTC-4.0: Updating the Gravitational-wave Transient Catalog with Observations from the First Part of the Fourth LIGO─Virgo─KAGRA Observing Run}",
      journal = {\apjl},
         year = 2026,
        month = jun,
       volume = {1004},
       number = {2},
          eid = {L22},
        pages = {L22},
          doi = {10.3847/2041-8213/ae2c74},
       adsurl = {https://ui.adsabs.harvard.edu/abs/2026ApJ..1004L..22A}
}

@ARTICLE{Schawinski:2015,
       author = {{Schawinski}, Kevin and {Koss}, Michael and {Berney}, Simon and {Sartori}, Lia F.},
        title = "{Active galactic nuclei flicker: an observational estimate of the duration of black hole growth phases of {\ensuremath{\sim}}{}10$^{5}$ yr}",
      journal = {\mnras},
         year = 2015,
        month = aug,
       volume = {451},
       number = {3},
        pages = {2517-2523},
          doi = {10.1093/mnras/stv1136},
archivePrefix = {arXiv},
       eprint = {1505.06733},
 primaryClass = {astro-ph.GA},
       adsurl = {https://ui.adsabs.harvard.edu/abs/2015MNRAS.451.2517S}
}

@ARTICLE{Paardekooper2006,
       author = {{Paardekooper}, S. -J. and {Mellema}, G.},
        title = "{Halting type I planet migration in non-isothermal disks}",
      journal = {\aap},
         year = 2006,
        month = nov,
       volume = {459},
       number = {1},
        pages = {L17-L20},
          doi = {10.1051/0004-6361:20066304},
archivePrefix = {arXiv},
       eprint = {astro-ph/0608658},
 primaryClass = {astro-ph},
       adsurl = {https://ui.adsabs.harvard.edu/abs/2006A&A...459L..17P}
}

@article{cresswell2008,
	author = {{Cresswell}, P. and {Nelson}, R. P.},
	journal = {A\&A},
	number = 2,
	pages = {677-690},
	title = {Three-dimensional simulations of multiple protoplanets embedded in a protostellar disc},
	volume = 482,
	year = 2008}

@ARTICLE{Laughlin_1994,
       author = {{Laughlin}, Gregory and {Bodenheimer}, Peter},
        title = "{Nonaxisymmetric Evolution in Protostellar Disks}",
      journal = {\apj},
         year = 1994,
        month = nov,
       volume = {436},
        pages = {335},
          doi = {10.1086/174909},
       adsurl = {https://ui.adsabs.harvard.edu/abs/1994ApJ...436..335L}
}

@ARTICLE{Ogihara_2007,
       author = {{Ogihara}, Masahiro and {Ida}, Shigeru and {Morbidelli}, Alessandro},
        title = "{Accretion of terrestrial planets from oligarchs in a turbulent disk}",
      journal = {\icarus},
         year = 2007,
        month = jun,
       volume = {188},
       number = {2},
        pages = {522-534},
          doi = {10.1016/j.icarus.2006.12.006},
archivePrefix = {arXiv},
       eprint = {astro-ph/0612619},
 primaryClass = {astro-ph},
       adsurl = {https://ui.adsabs.harvard.edu/abs/2007Icar..188..522O}
}

@ARTICLE{Stone17,
       author = {{Stone}, Nicholas C. and {Metzger}, Brian D. and {Haiman}, Zolt{\'a}n},
        title = "{Assisted inspirals of stellar mass black holes embedded in AGN discs: solving the `final au problem'}",
      journal = {\mnras},
         year = 2017,
        month = jan,
       volume = {464},
       number = {1},
        pages = {946-954},
          doi = {10.1093/mnras/stw2260},
archivePrefix = {arXiv},
       eprint = {1602.04226},
 primaryClass = {astro-ph.GA},
       adsurl = {https://ui.adsabs.harvard.edu/abs/2017MNRAS.464..946S}
}

@ARTICLE{Bartos17,
       author = {{Bartos}, Imre and {Kocsis}, Bence and {Haiman}, Zolt{\'a}n and {M{\'a}rka}, Szabolcs},
        title = "{Rapid and Bright Stellar-mass Binary Black Hole Mergers in Active Galactic Nuclei}",
      journal = {\apj},
         year = 2017,
        month = feb,
       volume = {835},
       number = {2},
          eid = {165},
        pages = {165},
          doi = {10.3847/1538-4357/835/2/165},
archivePrefix = {arXiv},
       eprint = {1602.03831},
 primaryClass = {astro-ph.HE},
       adsurl = {https://ui.adsabs.harvard.edu/abs/2017ApJ...835..165B}
}

@ARTICLE{McK22,
       author = {{McKernan}, B. and {Ford}, K.~E.~S. and {Callister}, T. and {Farr}, W.~M. and {O'Shaughnessy}, R. and {Smith}, R. and {Thrane}, E. and {Vajpeyi}, A.},
        title = "{LIGO-Virgo correlations between mass ratio and effective inspiral spin: testing the active galactic nuclei channel}",
      journal = {\mnras},
         year = 2022,
        month = aug,
       volume = {514},
       number = {3},
        pages = {3886-3893},
          doi = {10.1093/mnras/stac1570},
archivePrefix = {arXiv},
       eprint = {2107.07551},
 primaryClass = {astro-ph.HE},
       adsurl = {https://ui.adsabs.harvard.edu/abs/2022MNRAS.514.3886M}
}

@ARTICLE{Wang21,
       author = {{Wang}, Yi-Han and {McKernan}, Barry and {Ford}, Saavik and {Perna}, Rosalba and {Leigh}, Nathan W.~C. and {Mac Low}, Mordecai-Mark},
        title = "{Symmetry Breaking in Dynamical Encounters in the Disks of Active Galactic Nuclei}",
      journal = {\apjl},
         year = 2021,
        month = dec,
       volume = {923},
       number = {2},
          eid = {L23},
        pages = {L23},
          doi = {10.3847/2041-8213/ac400a},
archivePrefix = {arXiv},
       eprint = {2110.03698},
 primaryClass = {astro-ph.HE},
       adsurl = {https://ui.adsabs.harvard.edu/abs/2021ApJ...923L..23W}
}

@ARTICLE{Gerosa21,
       author = {{Gerosa}, Davide and {Fishbach}, Maya},
        title = "{Hierarchical mergers of stellar-mass black holes and their gravitational-wave signatures}",
      journal = {Nature Astronomy},
         year = 2021,
        month = jul,
       volume = {5},
        pages = {749-760},
          doi = {10.1038/s41550-021-01398-w},
archivePrefix = {arXiv},
       eprint = {2105.03439},
 primaryClass = {astro-ph.HE},
       adsurl = {https://ui.adsabs.harvard.edu/abs/2021NatAs...5..749G}
}

@ARTICLE{Mapelli21,
       author = {{Mapelli}, Michela and {Dall'Amico}, Marco and {Bouffanais}, Yann and {Giacobbo}, Nicola and {Arca Sedda}, Manuel and {Artale}, M. Celeste and {Ballone}, Alessandro and {Di Carlo}, Ugo N. and {Iorio}, Giuliano and {Santoliquido}, Filippo and {Torniamenti}, Stefano},
        title = "{Hierarchical black hole mergers in young, globular and nuclear star clusters: the effect of metallicity, spin and cluster properties}",
      journal = {\mnras},
         year = 2021,
        month = jul,
       volume = {505},
       number = {1},
        pages = {339-358},
          doi = {10.1093/mnras/stab1334},
archivePrefix = {arXiv},
       eprint = {2103.05016},
 primaryClass = {astro-ph.HE},
       adsurl = {https://ui.adsabs.harvard.edu/abs/2021MNRAS.505..339M}
}

@ARTICLE{McK20,
       author = {{McKernan}, B. and {Ford}, K.~E.~S. and {O'Shaugnessy}, R. and {Wysocki}, D.},
        title = "{Monte Carlo simulations of black hole mergers in AGN discs: Low {\ensuremath{\chi}}$_{eff}$ mergers and predictions for LIGO}",
      journal = {\mnras},
         year = 2020,
        month = may,
       volume = {494},
       number = {1},
        pages = {1203-1216},
          doi = {10.1093/mnras/staa740},
archivePrefix = {arXiv},
       eprint = {1907.04356},
 primaryClass = {astro-ph.HE},
       adsurl = {https://ui.adsabs.harvard.edu/abs/2020MNRAS.494.1203M}
}

@ARTICLE{Kremer20,
       author = {{Kremer}, Kyle and {Spera}, Mario and {Becker}, Devin and {Chatterjee}, Sourav and {Di Carlo}, Ugo N. and {Fragione}, Giacomo and {Rodriguez}, Carl L. and {Ye}, Claire S. and {Rasio}, Frederic A.},
        title = "{Populating the Upper Black Hole Mass Gap through Stellar Collisions in Young Star Clusters}",
      journal = {\apj},
         year = 2020,
        month = nov,
       volume = {903},
       number = {1},
          eid = {45},
        pages = {45},
          doi = {10.3847/1538-4357/abb945},
archivePrefix = {arXiv},
       eprint = {2006.10771},
 primaryClass = {astro-ph.HE},
       adsurl = {https://ui.adsabs.harvard.edu/abs/2020ApJ...903...45K}
}

@ARTICLE{Fragione19,
       author = {{Fragione}, Giacomo and {Grishin}, Evgeni and {Leigh}, Nathan W.~C. and {Perets}, Hagai B. and {Perna}, Rosalba},
        title = "{Black hole and neutron star mergers in galactic nuclei}",
      journal = {\mnras},
         year = 2019,
        month = sep,
       volume = {488},
       number = {1},
        pages = {47-63},
          doi = {10.1093/mnras/stz1651},
archivePrefix = {arXiv},
       eprint = {1811.10627},
 primaryClass = {astro-ph.GA},
       adsurl = {https://ui.adsabs.harvard.edu/abs/2019MNRAS.488...47F}
}

@ARTICLE{Rodriguez18,
       author = {{Rodriguez}, Carl L. and {Amaro-Seoane}, Pau and {Chatterjee}, Sourav and {Rasio}, Frederic A.},
        title = "{Post-Newtonian Dynamics in Dense Star Clusters: Highly Eccentric, Highly Spinning, and Repeated Binary Black Hole Mergers}",
      journal = {\prl},
         year = 2018,
        month = apr,
       volume = {120},
       number = {15},
          eid = {151101},
        pages = {151101},
          doi = {10.1103/PhysRevLett.120.151101},
archivePrefix = {arXiv},
       eprint = {1712.04937},
 primaryClass = {astro-ph.HE},
       adsurl = {https://ui.adsabs.harvard.edu/abs/2018PhRvL.120o1101R}
}

@ARTICLE{Antonini16,
       author = {{Antonini}, Fabio and {Rasio}, Frederic A.},
        title = "{Merging Black Hole Binaries in Galactic Nuclei: Implications for Advanced-LIGO Detections}",
      journal = {\apj},
         year = 2016,
        month = nov,
       volume = {831},
       number = {2},
          eid = {187},
        pages = {187},
          doi = {10.3847/0004-637X/831/2/187},
archivePrefix = {arXiv},
       eprint = {1606.04889},
 primaryClass = {astro-ph.HE},
       adsurl = {https://ui.adsabs.harvard.edu/abs/2016ApJ...831..187A}
}

@ARTICLE{Callister21,
       author = {{Callister}, Thomas A. and {Haster}, Carl-Johan and {Ng}, Ken K.~Y. and {Vitale}, Salvatore and {Farr}, Will M.},
        title = "{Who Ordered That? Unequal-mass Binary Black Hole Mergers Have Larger Effective Spins}",
      journal = {\apjl},
         year = 2021,
        month = nov,
       volume = {922},
       number = {1},
          eid = {L5},
        pages = {L5},
          doi = {10.3847/2041-8213/ac2ccc},
archivePrefix = {arXiv},
       eprint = {2106.00521},
 primaryClass = {astro-ph.HE},
       adsurl = {https://ui.adsabs.harvard.edu/abs/2021ApJ...922L...5C}
}

@ARTICLE{Adamcewicz23,
       author = {{Adamcewicz}, Christian and {Lasky}, Paul D. and {Thrane}, Eric},
        title = "{Evidence for a Correlation between Binary Black Hole Mass Ratio and Black Hole Spins}",
      journal = {\apj},
         year = 2023,
        month = nov,
       volume = {958},
       number = {1},
          eid = {13},
        pages = {13},
          doi = {10.3847/1538-4357/acf763},
archivePrefix = {arXiv},
       eprint = {2307.15278},
 primaryClass = {astro-ph.HE},
       adsurl = {https://ui.adsabs.harvard.edu/abs/2023ApJ...958...13A}
}

@ARTICLE{Santini23,
       author = {{Santini}, Alessandro and {Gerosa}, Davide and {Cotesta}, Roberto and {Berti}, Emanuele},
        title = "{Black-hole mergers in disklike environments could explain the observed q -{\ensuremath{\chi}}$_{eff}$ correlation}",
      journal = {\prd},
         year = 2023,
        month = oct,
       volume = {108},
       number = {8},
          eid = {083033},
        pages = {083033},
          doi = {10.1103/PhysRevD.108.083033},
archivePrefix = {arXiv},
       eprint = {2308.12998},
 primaryClass = {astro-ph.HE},
       adsurl = {https://ui.adsabs.harvard.edu/abs/2023PhRvD.108h3033S}
}

@ARTICLE{Fabj20,
       author = {{Fabj}, Gaia and {Nasim}, Syeda S. and {Caban}, Freddy and {Ford}, K.~E. Saavik and {McKernan}, Barry and {Bellovary}, Jillian M.},
        title = "{Aligning nuclear cluster orbits with an active galactic nucleus accretion disc}",
      journal = {\mnras},
         year = 2020,
        month = dec,
       volume = {499},
       number = {2},
        pages = {2608-2616},
          doi = {10.1093/mnras/staa3004},
archivePrefix = {arXiv},
       eprint = {2006.11229},
 primaryClass = {astro-ph.GA},
       adsurl = {https://ui.adsabs.harvard.edu/abs/2020MNRAS.499.2608F}
}

@ARTICLE{Nasim23,
       author = {{Nasim}, Syeda S. and {Fabj}, Gaia and {Caban}, Freddy and {Secunda}, Amy and {Ford}, K.~E. Saavik and {McKernan}, Barry and {Bellovary}, Jillian M. and {Leigh}, Nathan W.~C. and {Lyra}, Wladimir},
        title = "{Aligning Retrograde Nuclear Cluster Orbits with an Active Galactic Nucleus Accretion Disc}",
      journal = {\mnras},
         year = 2023,
        month = jul,
       volume = {522},
       number = {4},
        pages = {5393-5401},
          doi = {10.1093/mnras/stad1295},
archivePrefix = {arXiv},
       eprint = {2207.09540},
 primaryClass = {astro-ph.GA},
       adsurl = {https://ui.adsabs.harvard.edu/abs/2023MNRAS.522.5393N}
}

@ARTICLE{Generozov23,
       author = {{Generozov}, A. and {Perets}, H.~B.},
        title = "{Capture of stars into gaseous discs around massive black holes: alignment, circularization, and growth}",
      journal = {\mnras},
         year = 2023,
        month = jun,
       volume = {522},
       number = {2},
        pages = {1763-1778},
          doi = {10.1093/mnras/stad1016},
archivePrefix = {arXiv},
       eprint = {2212.11301},
 primaryClass = {astro-ph.GA},
       adsurl = {https://ui.adsabs.harvard.edu/abs/2023MNRAS.522.1763G}
}

@ARTICLE{Vaccaro23,
       author = {{Vaccaro}, Maria Paola and {Mapelli}, Michela and {P{\'e}rigois}, Carole and {Barone}, Dario and {Artale}, Maria Celeste and {Dall'Amico}, Marco and {Iorio}, Giuliano and {Torniamenti}, Stefano},
        title = "{Impact of gas hardening on the population properties of hierarchical black hole mergers in active galactic nucleus disks}",
      journal = {\aap},
         year = 2024,
        month = may,
       volume = {685},
          eid = {A51},
        pages = {A51},
          doi = {10.1051/0004-6361/202348509},
archivePrefix = {arXiv},
       eprint = {2311.18548},
 primaryClass = {astro-ph.HE},
       adsurl = {https://ui.adsabs.harvard.edu/abs/2024A&A...685A..51V}
}

@ARTICLE{Heger_2002,
       author = {{Heger}, A. and {Woosley}, S.~E.},
        title = "{The Nucleosynthetic Signature of Population III}",
      journal = {\apj},
         year = 2002,
        month = mar,
       volume = {567},
       number = {1},
        pages = {532-543},
          doi = {10.1086/338487},
archivePrefix = {arXiv},
       eprint = {astro-ph/0107037},
 primaryClass = {astro-ph},
       adsurl = {https://ui.adsabs.harvard.edu/abs/2002ApJ...567..532H}
}

@ARTICLE{Woosley_2007,
       author = {{Woosley}, S.~E. and {Blinnikov}, S. and {Heger}, Alexander},
        title = "{Pulsational pair instability as an explanation for the most luminous supernovae}",
      journal = {\nat},
         year = 2007,
        month = nov,
       volume = {450},
       number = {7168},
        pages = {390-392},
          doi = {10.1038/nature06333},
archivePrefix = {arXiv},
       eprint = {0710.3314},
 primaryClass = {astro-ph},
       adsurl = {https://ui.adsabs.harvard.edu/abs/2007Natur.450..390W}
}

@ARTICLE{Farmer_2019,
       author = {{Farmer}, R. and {Renzo}, M. and {de Mink}, S.~E. and {Marchant}, P. and {Justham}, S.},
        title = "{Mind the Gap: The Location of the Lower Edge of the Pair-instability Supernova Black Hole Mass Gap}",
      journal = {\apj},
         year = 2019,
        month = dec,
       volume = {887},
       number = {1},
          eid = {53},
        pages = {53},
          doi = {10.3847/1538-4357/ab518b},
archivePrefix = {arXiv},
       eprint = {1910.12874},
 primaryClass = {astro-ph.SR},
       adsurl = {https://ui.adsabs.harvard.edu/abs/2019ApJ...887...53F}
}

@ARTICLE{Woosley_2021,
       author = {{Woosley}, S.~E. and {Heger}, Alexander},
        title = "{The Pair-instability Mass Gap for Black Holes}",
      journal = {\apjl},
         year = 2021,
        month = may,
       volume = {912},
       number = {2},
          eid = {L31},
        pages = {L31},
          doi = {10.3847/2041-8213/abf2c4},
archivePrefix = {arXiv},
       eprint = {2103.07933},
 primaryClass = {astro-ph.SR},
       adsurl = {https://ui.adsabs.harvard.edu/abs/2021ApJ...912L..31W}
}

@ARTICLE{Tagawa21,
       author = {{Tagawa}, Hiromichi and {Haiman}, Zolt{\'a}n and {Bartos}, Imre and {Kocsis}, Bence and {Omukai}, Kazuyuki},
        title = "{Signatures of hierarchical mergers in black hole spin and mass distribution}",
      journal = {\mnras},
         year = 2021,
        month = nov,
       volume = {507},
       number = {3},
        pages = {3362-3380},
          doi = {10.1093/mnras/stab2315},
archivePrefix = {arXiv},
       eprint = {2104.09510},
 primaryClass = {astro-ph.HE},
       adsurl = {https://ui.adsabs.harvard.edu/abs/2021MNRAS.507.3362T}
}

@ARTICLE{Wang:2022,
       author = {{Wang}, Yuan-Zhu and {Li}, Yin-Jie and {Vink}, Jorick S. and {Fan}, Yi-Zhong and {Tang}, Shao-Peng and {Qin}, Ying and {Wei}, Da-Ming},
        title = "{Potential Subpopulations and Assembling Tendency of the Merging Black Holes}",
      journal = {\apjl},
         year = 2022,
        month = dec,
       volume = {941},
       number = {2},
          eid = {L39},
        pages = {L39},
          doi = {10.3847/2041-8213/aca89f},
archivePrefix = {arXiv},
       eprint = {2208.11871},
 primaryClass = {astro-ph.HE},
       adsurl = {https://ui.adsabs.harvard.edu/abs/2022ApJ...941L..39W}
}

@ARTICLE{Sadiq:2025,
       author = {{Sadiq}, Jam and {Dent}, Thomas and {Lorenzo-Medina}, Ana},
        title = "{Probing evolution in the black hole spectrum with gravitational waves catalogs}",
      journal = {\prd},
         year = 2025,
        month = oct,
       volume = {112},
       number = {8},
          eid = {083028},
        pages = {083028},
          doi = {10.1103/j9fq-nkl3},
archivePrefix = {arXiv},
       eprint = {2502.06451},
 primaryClass = {gr-qc},
       adsurl = {https://ui.adsabs.harvard.edu/abs/2025PhRvD.112h3028S}
}

@ARTICLE{Li:2025,
       author = {{Li}, Yin-Jie and {Wang}, Yuan-Zhu and {Tang}, Shao-Peng and {Chen}, Tong and {Fan}, Yi-Zhong},
        title = "{Revealing the {\ensuremath{\chi}}$_{eff}${\textendash}q Correlation among Coalescing Binary Black Holes and Tentative Evidence for AGN-driven Hierarchical Mergers}",
      journal = {\apj},
         year = 2025,
        month = jul,
       volume = {987},
       number = {1},
          eid = {65},
        pages = {65},
          doi = {10.3847/1538-4357/add535},
archivePrefix = {arXiv},
       eprint = {2501.09495},
 primaryClass = {astro-ph.HE},
       adsurl = {https://ui.adsabs.harvard.edu/abs/2025ApJ...987...65L}
}

@ARTICLE{Li:2024,
       author = {{Li}, Yin-Jie and {Wang}, Yuan-Zhu and {Tang}, Shao-Peng and {Fan}, Yi-Zhong},
        title = "{Resolving the Stellar-Collapse and Hierarchical-Merger Origins of the Coalescing Black Holes}",
      journal = {\prl},
         year = 2024,
        month = aug,
       volume = {133},
       number = {5},
          eid = {051401},
        pages = {051401},
          doi = {10.1103/PhysRevLett.133.051401},
archivePrefix = {arXiv},
       eprint = {2303.02973},
 primaryClass = {astro-ph.HE},
       adsurl = {https://ui.adsabs.harvard.edu/abs/2024PhRvL.133e1401L}
}

@ARTICLE{Guo:2024,
       author = {{Guo}, Wei-Hua and {Li}, Yin-Jie and {Wang}, Yuan-Zhu and {Shao}, Yong and {Wu}, Shi-Chao and {Zhu}, Tao and {Fan}, Yi-Zhong},
        title = "{The Heavier the Faster: A Subpopulation of Heavy, Rapidly Spinning and Quickly Evolving Binary Black Holes}",
      journal = {\apj},
         year = 2024,
        month = nov,
       volume = {975},
       number = {1},
          eid = {54},
        pages = {54},
          doi = {10.3847/1538-4357/ad758a},
archivePrefix = {arXiv},
       eprint = {2406.03257},
 primaryClass = {astro-ph.HE},
       adsurl = {https://ui.adsabs.harvard.edu/abs/2024ApJ...975...54G}
}

@ARTICLE{Antonini:2025,
       author = {{Antonini}, Fabio and {Romero-Shaw}, Isobel M. and {Callister}, Thomas},
        title = "{Star Cluster Population of High Mass Black Hole Mergers in Gravitational Wave Data}",
      journal = {\prl},
         year = 2025,
        month = jan,
       volume = {134},
       number = {1},
          eid = {011401},
        pages = {011401},
          doi = {10.1103/PhysRevLett.134.011401},
archivePrefix = {arXiv},
       eprint = {2406.19044},
 primaryClass = {astro-ph.HE},
       adsurl = {https://ui.adsabs.harvard.edu/abs/2025PhRvL.134a1401A}
}

@ARTICLE{Godfrey:2023,
       author = {{Godfrey}, Jaxen and {Edelman}, Bruce and {Farr}, Ben},
        title = "{Cosmic Cousins: Identification of a Subpopulation of Binary Black Holes Consistent with Isolated Binary Evolution}",
      journal = {arXiv e-prints},
         year = 2023,
        month = apr,
          eid = {arXiv:2304.01288},
        pages = {arXiv:2304.01288},
          doi = {10.48550/arXiv.2304.01288},
archivePrefix = {arXiv},
       eprint = {2304.01288},
 primaryClass = {astro-ph.HE},
       adsurl = {https://ui.adsabs.harvard.edu/abs/2023arXiv230401288G}
}

@ARTICLE{Mould:2022,
       author = {{Mould}, Matthew and {Gerosa}, Davide},
        title = "{Gravitational-wave population inference at past time infinity}",
      journal = {\prd},
         year = 2022,
        month = jan,
       volume = {105},
       number = {2},
          eid = {024076},
        pages = {024076},
          doi = {10.1103/PhysRevD.105.024076},
archivePrefix = {arXiv},
       eprint = {2110.05507},
 primaryClass = {astro-ph.HE},
       adsurl = {https://ui.adsabs.harvard.edu/abs/2022PhRvD.105b4076M}
}

@ARTICLE{Sandor_2011,
       author = {{S{\'a}ndor}, Zsolt and {Lyra}, Wladimir and {Dullemond}, Cornelis P.},
        title = "{Formation of Planetary Cores at Type I Migration Traps}",
      journal = {\apjl},
         year = 2011,
        month = feb,
       volume = {728},
       number = {1},
          eid = {L9},
        pages = {L9},
          doi = {10.1088/2041-8205/728/1/L9},
archivePrefix = {arXiv},
       eprint = {1101.0942},
 primaryClass = {astro-ph.EP},
       adsurl = {https://ui.adsabs.harvard.edu/abs/2011ApJ...728L...9S}
}

@ARTICLE{Secunda_2019,
       author = {{Secunda}, Amy and {Bellovary}, Jillian and {Mac Low}, Mordecai-Mark and {Ford}, K.~E. Saavik and {McKernan}, Barry and {Leigh}, Nathan W.~C. and {Lyra}, Wladimir and {S{\'a}ndor}, Zsolt},
        title = "{Orbital Migration of Interacting Stellar Mass Black Holes in Disks around Supermassive Black Holes}",
      journal = {\apj},
         year = 2019,
        month = jun,
       volume = {878},
       number = {2},
          eid = {85},
        pages = {85},
          doi = {10.3847/1538-4357/ab20ca},
archivePrefix = {arXiv},
       eprint = {1807.02859},
 primaryClass = {astro-ph.HE},
       adsurl = {https://ui.adsabs.harvard.edu/abs/2019ApJ...878...85S}
}

@ARTICLE{Secunda_2020,
       author = {{Secunda}, Amy and {Bellovary}, Jillian and {Mac Low}, Mordecai-Mark and {Ford}, K.~E. Saavik and {McKernan}, Barry and {Leigh}, Nathan W.~C. and {Lyra}, Wladimir and {S{\'a}ndor}, Zsolt and {Adorno}, Jose I.},
        title = "{Orbital Migration of Interacting Stellar Mass Black Holes in Disks around Supermassive Black Holes. II. Spins and Incoming Objects}",
      journal = {\apj},
         year = 2020,
        month = nov,
       volume = {903},
       number = {2},
          eid = {133},
        pages = {133},
          doi = {10.3847/1538-4357/abbc1d},
archivePrefix = {arXiv},
       eprint = {2004.11936},
 primaryClass = {astro-ph.HE},
       adsurl = {https://ui.adsabs.harvard.edu/abs/2020ApJ...903..133S}
}

@ARTICLE{Horn_2012,
       author = {{Horn}, Brandon and {Lyra}, Wladimir and {Mac Low}, Mordecai-Mark and {S{\'a}ndor}, Zsolt},
        title = "{Orbital Migration of Interacting Low-mass Planets in Evolutionary Radiative Turbulent Models}",
      journal = {\apj},
         year = 2012,
        month = may,
       volume = {750},
       number = {1},
          eid = {34},
        pages = {34},
          doi = {10.1088/0004-637X/750/1/34},
archivePrefix = {arXiv},
       eprint = {1202.1868},
 primaryClass = {astro-ph.EP},
       adsurl = {https://ui.adsabs.harvard.edu/abs/2012ApJ...750...34H}
}

@ARTICLE{Bellovary_2016,
       author = {{Bellovary}, Jillian M. and {Mac Low}, Mordecai-Mark and {McKernan}, Barry and {Ford}, K.~E. Saavik},
        title = "{Migration Traps in Disks around Supermassive Black Holes}",
      journal = {\apjl},
         year = 2016,
        month = mar,
       volume = {819},
       number = {2},
          eid = {L17},
        pages = {L17},
          doi = {10.3847/2041-8205/819/2/L17},
archivePrefix = {arXiv},
       eprint = {1511.00005},
 primaryClass = {astro-ph.GA},
       adsurl = {https://ui.adsabs.harvard.edu/abs/2016ApJ...819L..17B}
}

@ARTICLE{Schneider_2023,
       author = {{Schneider}, Fabian R.~N. and {Podsiadlowski}, Philipp and {Laplace}, Eva},
        title = "{Bimodal Black Hole Mass Distribution and Chirp Masses of Binary Black Hole Mergers}",
      journal = {\apjl},
         year = 2023,
        month = jun,
       volume = {950},
       number = {2},
          eid = {L9},
        pages = {L9},
          doi = {10.3847/2041-8213/acd77a},
archivePrefix = {arXiv},
       eprint = {2305.02380},
 primaryClass = {astro-ph.HE},
       adsurl = {https://ui.adsabs.harvard.edu/abs/2023ApJ...950L...9S}
}

@ARTICLE{Sirko&Goodman_2003,
       author = {{Sirko}, Edwin and {Goodman}, Jeremy},
        title = "{Spectral energy distributions of marginally self-gravitating quasi-stellar object discs}",
      journal = {\mnras},
         year = 2003,
        month = may,
       volume = {341},
       number = {2},
        pages = {501-508},
          doi = {10.1046/j.1365-8711.2003.06431.x},
archivePrefix = {arXiv},
       eprint = {astro-ph/0209469},
 primaryClass = {astro-ph},
       adsurl = {https://ui.adsabs.harvard.edu/abs/2003MNRAS.341..501S}
}

@ARTICLE{Paardekooper_2010,
       author = {{Paardekooper}, S. -J. and {Baruteau}, C. and {Crida}, A. and {Kley}, W.},
        title = "{A torque formula for non-isothermal type I planetary migration - I. Unsaturated horseshoe drag}",
      journal = {\mnras},
         year = 2010,
        month = jan,
       volume = {401},
       number = {3},
        pages = {1950-1964},
          doi = {10.1111/j.1365-2966.2009.15782.x},
archivePrefix = {arXiv},
       eprint = {0909.4552},
 primaryClass = {astro-ph.EP},
       adsurl = {https://ui.adsabs.harvard.edu/abs/2010MNRAS.401.1950P}
}

@ARTICLE{Salpeter_1955,
       author = {{Salpeter}, Edwin E.},
        title = "{The Luminosity Function and Stellar Evolution.}",
      journal = {\apj},
         year = 1955,
        month = jan,
       volume = {121},
        pages = {161},
          doi = {10.1086/145971},
       adsurl = {https://ui.adsabs.harvard.edu/abs/1955ApJ...121..161S}
}

@ARTICLE{Deck:2013,
       author = {{Deck}, Katherine M. and {Payne}, Matthew and {Holman}, Matthew J.},
        title = "{First-order Resonance Overlap and the Stability of Close Two-planet Systems}",
      journal = {\apj},
         year = 2013,
        month = sep,
       volume = {774},
       number = {2},
          eid = {129},
        pages = {129},
          doi = {10.1088/0004-637X/774/2/129},
archivePrefix = {arXiv},
       eprint = {1307.8119},
 primaryClass = {astro-ph.EP},
       adsurl = {https://ui.adsabs.harvard.edu/abs/2013ApJ...774..129D}
}

@article{Mckernan_2018,
       author = {{McKernan}, Barry and {Ford}, K.~E. Saavik and {Bellovary}, J. and {Leigh}, N.~W.~C. and {Haiman}, Z. and {Kocsis}, B. and {Lyra}, W. and {Mac Low}, M.-M. and {Metzger}, B. and {O'Dowd}, M. and et al.},
        title = "{Constraining Stellar-mass Black Hole Mergers in AGN Disks Detectable with LIGO}",
      journal = {\apj},
         year = 2018,
        month = oct,
       volume = {866},
       number = {1},
          eid = {66},
        pages = {66},
          doi = {10.3847/1538-4357/aadae5},
archivePrefix = {arXiv},
       eprint = {1702.07818},
 primaryClass = {astro-ph.HE},
       adsurl = {https://ui.adsabs.harvard.edu/abs/2018ApJ...866...66M}
}

@ARTICLE{Lin:1986,
       author = {{Lin}, D.~N.~C. and {Papaloizou}, John},
        title = "{On the Tidal Interaction between Protoplanets and the Protoplanetary Disk. III. Orbital Migration of Protoplanets}",
      journal = {\apj},
         year = 1986,
        month = oct,
       volume = {309},
        pages = {846},
          doi = {10.1086/164653},
       adsurl = {https://ui.adsabs.harvard.edu/abs/1986ApJ...309..846L}
}

@ARTICLE{Rowan2023,
       author = {{Rowan}, Connar and {Boekholt}, Tjarda and {Kocsis}, Bence and {Haiman}, Zolt{\'a}n},
        title = "{Black hole binary formation in AGN discs: from isolation to merger}",
      journal = {\mnras},
         year = 2023,
        month = sep,
       volume = {524},
       number = {2},
        pages = {2770-2796},
          doi = {10.1093/mnras/stad1926},
archivePrefix = {arXiv},
       eprint = {2212.06133},
 primaryClass = {astro-ph.GA},
       adsurl = {https://ui.adsabs.harvard.edu/abs/2023MNRAS.524.2770R}
}

@ARTICLE{Qian2024,
       author = {{Qian}, Kecheng and {Li}, Jiaru and {Lai}, Dong},
        title = "{Dynamical Friction Models for Black Hole Binary Formation in Active Galactic Nucleus Disks}",
      journal = {\apj},
         year = 2024,
        month = feb,
       volume = {962},
       number = {2},
          eid = {143},
        pages = {143},
          doi = {10.3847/1538-4357/ad1b53},
archivePrefix = {arXiv},
       eprint = {2310.12208},
 primaryClass = {astro-ph.HE},
       adsurl = {https://ui.adsabs.harvard.edu/abs/2024ApJ...962..143Q}
}

@ARTICLE{Whitehead2024,
       author = {{Whitehead}, Henry and {Rowan}, Connar and {Boekholt}, Tjarda and {Kocsis}, Bence},
        title = "{Gas assisted Binary Black Hole formation in AGN discs}",
      journal = {\mnras},
         year = 2024,
        month = jun,
          doi = {10.1093/mnras/stae1430},
archivePrefix = {arXiv},
       eprint = {2309.11561},
 primaryClass = {astro-ph.GA},
       adsurl = {https://ui.adsabs.harvard.edu/abs/2024MNRAS.tmp.1420W}
}

@ARTICLE{DeLaurentiis2023,
       author = {{DeLaurentiis}, Stanislav and {Epstein-Martin}, Marguerite and {Haiman}, Zolt{\'a}n},
        title = "{Gas dynamical friction as a binary formation mechanism in AGN discs}",
      journal = {\mnras},
         year = 2023,
        month = jul,
       volume = {523},
       number = {1},
        pages = {1126-1139},
          doi = {10.1093/mnras/stad1412},
archivePrefix = {arXiv},
       eprint = {2212.02650},
 primaryClass = {astro-ph.HE},
       adsurl = {https://ui.adsabs.harvard.edu/abs/2023MNRAS.523.1126D}
}

@ARTICLE{Zwart2002,
       author = {{Portegies Zwart}, Simon F. and {McMillan}, Stephen L.~W.},
        title = "{The Runaway Growth of Intermediate-Mass Black Holes in Dense Star Clusters}",
      journal = {\apj},
         year = 2002,
        month = sep,
       volume = {576},
       number = {2},
        pages = {899-907},
          doi = {10.1086/341798},
archivePrefix = {arXiv},
       eprint = {astro-ph/0201055},
 primaryClass = {astro-ph},
       adsurl = {https://ui.adsabs.harvard.edu/abs/2002ApJ...576..899P}
}

@ARTICLE{Miller2002,
       author = {{Miller}, M. Coleman and {Hamilton}, Douglas P.},
        title = "{Production of intermediate-mass black holes in globular clusters}",
      journal = {\mnras},
         year = 2002,
        month = feb,
       volume = {330},
       number = {1},
        pages = {232-240},
          doi = {10.1046/j.1365-8711.2002.05112.x},
archivePrefix = {arXiv},
       eprint = {astro-ph/0106188},
 primaryClass = {astro-ph},
       adsurl = {https://ui.adsabs.harvard.edu/abs/2002MNRAS.330..232C}
}

@ARTICLE{Carr2021,
       author = {{Carr}, Bernard and {Kohri}, Kazunori and {Sendouda}, Yuuiti and {Yokoyama}, Jun'ichi},
        title = "{Constraints on primordial black holes}",
      journal = {Reports on Progress in Physics},
         year = 2021,
        month = nov,
       volume = {84},
       number = {11},
          eid = {116902},
        pages = {116902},
          doi = {10.1088/1361-6633/ac1e31},
archivePrefix = {arXiv},
       eprint = {2002.12778},
 primaryClass = {astro-ph.CO},
       adsurl = {https://ui.adsabs.harvard.edu/abs/2021RPPh...84k6902C}
}







\end{document}